%% file: main.tex
\documentclass{elsarticle}
\usepackage{lineno}
\modulolinenumbers[1]

\usepackage{color}

\usepackage{framed,multirow}
\usepackage[letterpaper,top=1in,bottom=1in,left=1in,right=1in]{geometry}
\usepackage{amssymb}
\usepackage{latexsym}
\usepackage{arydshln}
\usepackage{blkarray}

\usepackage{url}
\usepackage{xcolor}
\definecolor{newcolor}{rgb}{.8,.349,.1}

\journal{Journal of Computational Physics}

\definecolor{mygreen}{HTML}{00bf7c}

\usepackage{amsmath}
\usepackage{subcaption}

\usepackage{xspace}
\usepackage{multirow}

\makeatletter
\DeclareRobustCommand\onedot{\futurelet\@let@token\@onedot}
\def\@onedot{\ifx\@let@token.\else.\null\fi\xspace}

\def\eg{e.g\onedot} 
\def\ie{i.e\onedot} 
 
\def\etc{etc\onedot}

\usepackage{scalerel,stackengine}
\stackMath

\include{commands}

\newcommand*\patchAmsMathEnvironmentForLineno[1]{%
   \expandafter\let\csname old#1\expandafter\endcsname\csname #1\endcsname
   \expandafter\let\csname oldend#1\expandafter\endcsname\csname end#1\endcsname
   \renewenvironment{#1}%
      {\linenomath\csname old#1\endcsname}%
      {\csname oldend#1\endcsname\endlinenomath}}%
\newcommand*\patchBothAmsMathEnvironmentsForLineno[1]{%
   \patchAmsMathEnvironmentForLineno{#1}%
   \patchAmsMathEnvironmentForLineno{#1*}}%
\AtBeginDocument{%
\patchBothAmsMathEnvironmentsForLineno{equation}%
\patchBothAmsMathEnvironmentsForLineno{align}%
\patchBothAmsMathEnvironmentsForLineno{flalign}%
\patchBothAmsMathEnvironmentsForLineno{alignat}%
\patchBothAmsMathEnvironmentsForLineno{gather}%
\patchBothAmsMathEnvironmentsForLineno{multline}%
}

\usepackage{tikz,environ}
\usetikzlibrary{decorations.pathreplacing,calligraphy}
\NewEnviron{dpmatrix}{\setbox0=\hbox{$\,\begin{matrix}\BODY\end{matrix}\,$}%
	\setbox2=\hbox{%
		\begin{tikzpicture}
		\draw [dashed, thick, decorate, decoration={calligraphic straight parenthesis, amplitude=1pt}] (0,\botdim)--(0,\topdim);
		\copy0
		\draw [dashed, thick, decorate, decoration={calligraphic straight parenthesis, amplitude=1pt}] (0,\topdim)--(0,\botdim);
		\end{tikzpicture}}
	\vcenter{\hbox{\copy2}}}
\def\topdim{\dimexpr+\ht0+.5\ht\strutbox-.5\dp\strutbox-3pt\relax}
\def\botdim{\dimexpr-\ht0+.5\ht\strutbox-.5\dp\strutbox+3pt\relax}

\makeatother

\begin{document}
	
	
	\begin{frontmatter}
		
		\title{Coercing Machine Learning to Output Physically Accurate Results}

\author{Zhenglin  Geng}
\author{Daniel Johnson}
\author{Ronald Fedkiw}
		
\address{Stanford University, 353 Jane Stanford Way, Gates Computer Science, Stanford, CA, 94305, United States}
		
		
\begin{abstract}
Many machine/deep learning artificial neural networks are trained to simply be interpolation functions that map input variables to output values interpolated from the training data in a linear/nonlinear fashion.
Even when the input/output pairs of the training data are physically accurate (e.g. the results of an experiment or numerical simulation), interpolated quantities can deviate quite far from being physically accurate.
Although one could project the output of a network into a physically feasible region, such a postprocess is not captured by the energy function minimized when training the network; thus, the final projected result could incorrectly deviate quite far from the training data.
We propose folding any such projection or postprocess directly into the network so that the final result is correctly compared to the training data by the energy function.
Although we propose a general approach, we illustrate its efficacy on a specific convolutional neural network that takes in human pose parameters (joint rotations) and outputs a prediction of vertex positions representing a triangulated cloth mesh.
While the original network outputs vertex positions with erroneously high stretching and compression energies, 
the new network trained with our physics “prior” remedies these issues producing highly improved results.
\end{abstract}
		
		
\end{frontmatter}
	\input{section1}
	\input{section2}
	\input{section3}
	\input{section4}
	\input{section5}
	\input{section6}
	\input{conclusion}
	\input{acknowledgement}
	\input{appendix}
	\bibliographystyle{model1-num-names}
	\bibliography{main}
\end{document}

%% file: commands.tex
\newcommand\reallywidehat[1]{%
	\savestack{\tmpbox}{\stretchto{%
			\scaleto{%
				\scalerel*[\widthof{\ensuremath{#1}}]{\kern-.6pt\bigwedge\kern-.6pt}%
				{\rule[-\textheight/2]{1ex}{\textheight}}
			}{\textheight}%
		}{0.5ex}}%
	\stackon[1pt]{#1}{\tmpbox}%
}

	\newcommand{\B}{\mathcal{B}}
	\newcommand{\C}{\mathcal{C}}
	\newcommand{\net}{\hat{r}}
	\newcommand{\skin}{\Tilde{r}}
	\renewcommand{\sim}{r}
	
	\newcommand{\lm}{l_e^{\text{max}}}
	\newcommand{\opt}{r^*}
	\newcommand{\E}{\mathcal{E}}
	\newcommand{\K}{\mathcal{K}}
	
	\newcommand{\xo}{x}
	\newcommand{\yo}{y}
	\newcommand{\z}{z}
	\newcommand{\So}{S}
	\newcommand{\Zo}{Z}
	\newcommand{\Et}{E}
	\newcommand{\pd}[2]{\frac{\partial #1}{\partial #2}}

	\newcommand{\stiffnessfigwidth}{0.19}
	\newcommand{\soo}{s_0^0}
	\newcommand{\sol}{s_0^1{}}
	\newcommand{\seo}{s_e^0}
	\newcommand{\sel}{s_e^1{}}
	\newcommand{\zoo}{z_0^0}
	\newcommand{\zol}{z_0^1{}}
	\newcommand{\zeo}{z_e^0}
	\newcommand{\zel}{z_e^1{}}

	\newcommand{\coo}{c_0^0}
	\newcommand{\col}{c_0^1}
	\newcommand{\ceo}{c_e^0}
	\newcommand{\cel}{c_e^1}
	\newcommand{\br}{b_r}
	\newcommand{\bt}{b_t}
	\newcommand{\doo}{d_0^0}
	\newcommand{\dol}{d_0^1{}}
	\newcommand{\deo}{d_e^0}
	\newcommand{\del}{d_e^1}
    \newcommand{\dashed}[1]{
		\setbox0=\hbox{$\hspace{1mm}\begin{matrix} #1 \end{matrix}\hspace{1mm}$}
		\noindent\vcenter{\vspace{1mm}\hbox{
		\begin{tikzpicture}
		\draw [dashed, thick, decorate, decoration={calligraphic straight parenthesis, amplitude=1pt}, anchor=west] (0,\botdim)--(0,\topdim);
		\copy0
		\draw [dashed, thick, decorate, decoration={calligraphic straight parenthesis, amplitude=1pt}, anchor=west] (0,\topdim)--(0,\botdim);
		\end{tikzpicture}
		}
		\vspace{1mm}}\hspace{\wd0}
		
	}

%% file: section1.tex
\section{Introduction}
Many aspects of physical problems are not well understood, and various modeling approximations have helped to make progress; however, many problems remain difficult, whether it be turbulence in fluid flows, surface tension and two-phase flows, coupling between adhesion and cohesion for contact angles, parameters for solid constitutive modeling, the conditions under which materials fracture, etc.
In all of these aforementioned examples, although physical experiments facilitate data generation, it is often unclear how to utilize this data in order to obtain models and parameters for use in numerical simulations.
Notably, the recent attention given to machine/deep learning stems from the ability to simply annotate data in various ways and subsequently train networks to interpolate from this data without requiring a full understanding or explicit modeling of the underlying system.
Of course, ignoring the physics may lead to wildly physically inaccurate results, even though those results might otherwise naively seem like valid interpolations.
In addition, such errors are often exacerbated by the use of sparse data representing a physically valid low-dimensional manifold in an otherwise high-dimensional space.
Since a number of authors have begun to consider the use of machine/deep learning for problems in traditional computational physics, see e.g. \cite{ling2016machine,parish2016paradigm,raissi2017machine,raissi2018hidden,tripathy2018deep,sirignano2018dgm,qi2019computing,chang2019identification,raissi2019physics,yeo2019deep,zhu2019physics,gibou2019sharp}, we are motivated to consider methodologies that constrain the interpolatory results of a network to be contained within a physically admissible region.
Quite recently, \cite{stinis2019enforcing} proposed adding physical constraints to generative adversarial networks (GANs) also considering projection as we do, while stressing the interplay between scientific computing and machine learning; we refer the interested reader to their work for even more motivation for such approaches.
	

Generally speaking, networks can be used to interpolate a function $f$ from known training pairs/examples $(x_T,y_T)$ with $y_T=f(x_T)$. The network approximation $\hat{f}(w,x)$ depends on parameters $w$ that specify the network so that $y_T\approx \hat{f}(w, x_T)$ for all $(x_T,y_T)$. Suitable parameters $w$ are typically found by minimizing an energy function of the form $\sum_T \lVert y_T - \hat{f}(w,x_T) \rVert$ with respect to $w$. The network architecture, \ie the form of $\hat{f}$, and the subsequent energy minimization are both extremely important for obtaining desirable results. For example, if $\hat{f}$ lacks the expressiveness to capture variability in the training data, \ie underfitting, there will be large errors in $\hat{f}(w, x_T)$ when compared to $y_T$. On the other hand, although one could create a network with many degrees of freedom in order to capture $y_T=\hat{f}(w,x_T)$ as accurately as desired, even exactly, $\hat{f}(w,x)$ could oscillate wildly and inaccurately when $x$ is not equal to $x_T$, \ie overfitting. See, \eg \cite{goodfellow2016deep,hastie2005elements,andrew2019lecture,li2019lecture}. One needs to take great care when designing the network architecture in order to avoid underfitting while still allowing for enough regularization to also avoid overfitting. Likewise, the form of the energy function and nature of the numerical optimization techniques also need careful consideration.
Some of the most popular methods include variants of BFGS \cite{broyden1970convergence,liu1989limited,dean2012large} and a number of methods based on gradient descent \cite{robbins1951stochastic,wright2015coordinate} (see also \cite{bao2018improved} and the references therein) or interpreting gradient descent as a numerical approximation to an ordinary differential equation to be solved via various approaches motivated by order of accuracy \cite{polyak1964some,nesterov1983method} and adaptive time-stepping \cite{duchi2011adaptive,zeiler2012adadelta,tieleman2012lecture,kingma2014adam,dozat2016incorporating}.

Devising a network architecture with enough representative capability to alleviate underfitting while still being amenable to the regularization required to avoid overfitting, and subsequently applying numerical optimization techniques to an adequately designed energy in order to find reasonable parameters $w$ is a quite difficult and mostly experimental endeavour. Thus, much of the progress made by the community emanates from the laborious creation of data sets that many researchers can consider in order to design network architectures and find suitable parameters $w$, see \eg \cite{imagenet_cvpr09}. This is typically driven by a community (rather than an individual or group) effort, and state-of-the-art results are often obtained incrementally by leveraging the works of others. Following this methodology, we choose an existing network and add a postprocess that projects the network's output to be physically admissible/feasible/accurate (see Sections~\ref{sec:inext_post} and \ref{sec:buck_post}). 
Importantly, any such postprocess needs to be robust enough to handle the potentially wildly physically inaccurate output of an interpolatory network.
Additionally, we incorporate this postprocess into the network itself by modifying the energy to be minimized to use the results of the postprocess instead of the network output (see Sections~\ref{sec:inext_prior} and \ref{sec:buck_prior}).
This approach requires that any such postprocess be differentiable enough to be embedded into the numerical optimization.
We demonstrate the efficacy of our approach on the convolutional neural network from \cite{jin2018pixel} that predicts a cloth shape from joint angles, showing that our procedure not only produces a more physically accurate result but also matches the ground truth as well as the original network.


Section~\ref{sec:data_driven_cloth} presents the details of the convolutional neural network from \cite{jin2018pixel}, and Section~\ref{sec:inext_post} describes how we postprocess the output of that network to make it more physically accurate.
Those more physically accurate results may deviate quite far from both the training data and the ground truth.
So, in Section~\ref{sec:inext_prior}, we incorporate the postprocess into the network itself as a so-called ”prior”, obtaining results that are not only physically accurate but also well match both the training data and the ground truth.
Sections~\ref{sec:inext_post} and \ref{sec:inext_prior} only consider over-stretched material and do not capture buckling phenomena, thus in sections~\ref{sec:buck_post} and \ref{sec:buck_prior} we duplicate the considerations of Section~\ref{sec:inext_post} and \ref{sec:inext_prior} for a more complex postprocess that also considers buckling.   


Regarding related works, our proposal to embed a second order cone program (Section~\ref{sec:inext_prior})  and a quasistaic physics simulation (Section~\ref{sec:buck_prior}) into a trained network appears to be novel to the best of our knowledge. \cite{amos2017optnet} did add a convex optimization layer, but they only considered a quadratic program with linear constraints, which is a subset of second order cone programming \cite{boyd2004convex}. Furthermore, they implemented a dense solver using a primal-dual method without using a faster solver enabled by second order cone programming. They only demonstrated examples with hundreds of variables whereas we considered around 10,000. Recently, \cite{akshay2019} proposed a method that differentiates through a cone program with generality; however, they assume invertibility whereas we show in Appendix A that our case has a null space. In addition, they obtain a non-symmetric system whereas we make our system symmetric in order to use a fast solver. This is quite important because even with a fast solver, the second order cone program approach in Section~\ref{sec:inext_post} and \ref{sec:inext_prior} is $40$ to $50$ times slower than the quasistatic approach in Section~\ref{sec:buck_post} and \ref{sec:buck_prior}. Regarding the quasistatic case, the derivatives are more straight-forward, although care must be taken for poor conditioning and inversion; however, this has all been previously addressed some time ago, see \cite{teran2003finite,irving2004invertible,sifakis2005automatic} which require only minor modifications (as discussed in \cite{cong2016art,bao2019high}) for our purposes.

%% file: section2.tex
	\section{Data-Driven Network for Cloth}
	\label{sec:data_driven_cloth}
	
\begin{figure}[t]
		\centering
		\begin{subfigure}[b]{0.19\linewidth}
			\includegraphics[width=\linewidth]{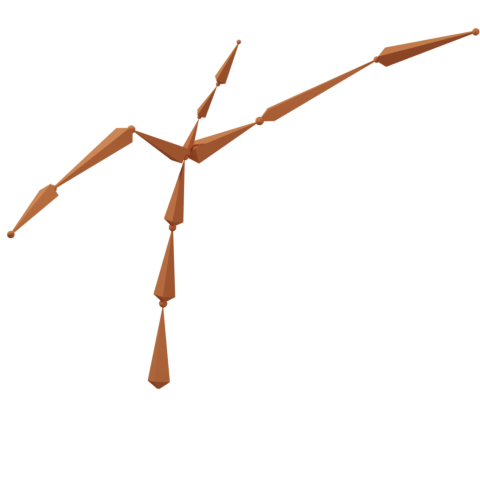}
			\caption{}
			\label{fig:skeleton}
		\end{subfigure}
		\begin{subfigure}[b]{0.19\linewidth}
			\includegraphics[width=\linewidth]{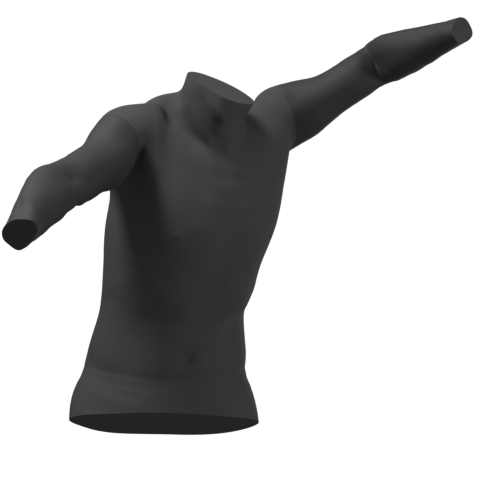}
			\caption{}
			\label{fig:body}
		\end{subfigure}
		\begin{subfigure}[b]{0.19\linewidth}
			\includegraphics[width=\linewidth]{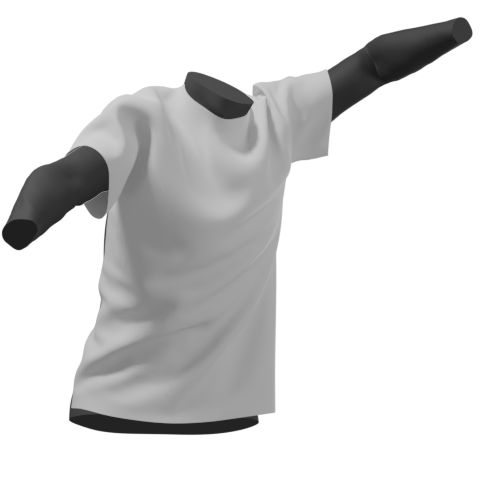}
			\caption{}
			\label{fig:clothed_body}
		\end{subfigure}
		\begin{subfigure}[b]{0.19\linewidth}
			\includegraphics[width=\linewidth]{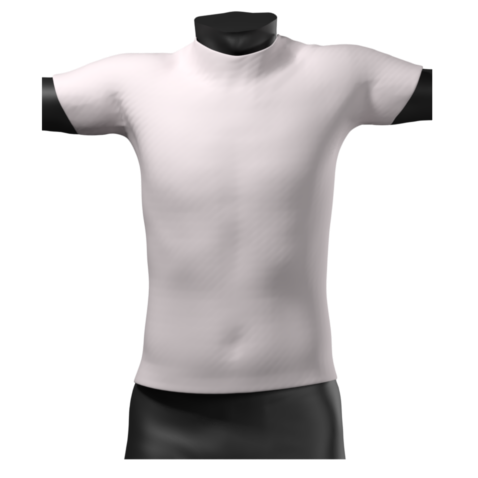}
			\caption{}
			\label{fig:shrink_wrap_neutral}
		\end{subfigure}
		\begin{subfigure}[b]{0.19\linewidth}
			\includegraphics[width=\linewidth]{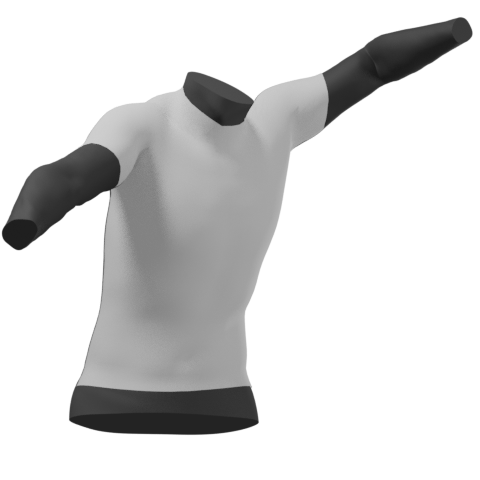}
			\caption{}
			\label{fig:shrink_wrap_posed}
		\end{subfigure}
		\caption{(a) A body pose specified by joint angles. (b)  A triangulated surface representing the body exterior for the pose shown in (a). (c) A cloth simulation detecting and processing collisions with the body geometry shown in (b). (d) Shrink-wrapped cloth on the neutral pose. (e) Shrink-wrapped cloth vertices follow their parent triangles as the body surface $\B$ deforms. 
		}
		\label{fig:preprocess}
	\end{figure}

	An articulated rigid body skeleton is posed by specifying a set of joint angles $\theta$, where it is assumed that the root node (\eg the pelvis) is fixed.  
	See Figure~\ref{fig:skeleton}. Then, a procedural skinning algorithm is used to form a triangulated surface $\B$ representing the exterior surface of the body in a manner consistent with the pose $\theta$. See Figure~\ref{fig:body}.
	There are a wide variety of approaches for skinning a body triangulated surface $\B$ from joint angles $\theta$, ranging from error-prone estimates to those with more bio-mechanical accuracy, see \eg \cite{Magnenat-Thalmann:1989,teran2003finite,Blemker2005,Kavan:2007,Stavness:2014,Le:2016,Lee:2019,Le:2019}.
	The body triangulated surface $\B$ is converted to a level set representation, see \eg \cite{osher2006level}, which is used to detect and process collisions with the triangulated surface $\C$ representing the cloth, see \eg \cite{Bridson:2003}.  See Figure~\ref{fig:clothed_body}.  
	Although any method could be used to simulate the cloth triangulated surface $\C$ with elasticity and bending, the authors of \cite{jin2018pixel} used methods derived from \cite{bridson2002robust,selle2008robust}. 
	As is typical, they carried out a number of numerical simulations to create a large data set of corresponding pairs $(\theta,\C)$;  
	a representative subset of those pairs, $(\theta_T,\C_T)$, was chosen as the training set for the network, and the remaining pairs were used to ascertain the predictive capabilities of any such network.  
	Notably, after finding suitable parameters $w$ for the network, the entire set of data (including the training pairs) may be discarded.
	
\begin{figure}[b]
	\centering
	\begin{subfigure}[b]{0.21\linewidth}
		\includegraphics[width=\linewidth]{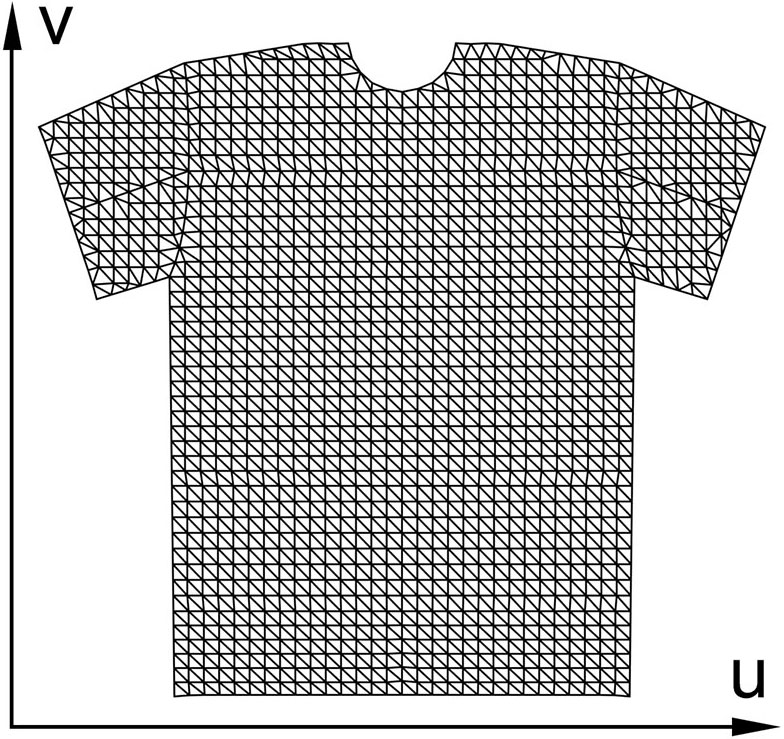}
		\caption{}
		\label{fig:front_flat_mesh}
	\end{subfigure}
	\begin{subfigure}[b]{0.245\linewidth}
		\includegraphics[width=\linewidth]{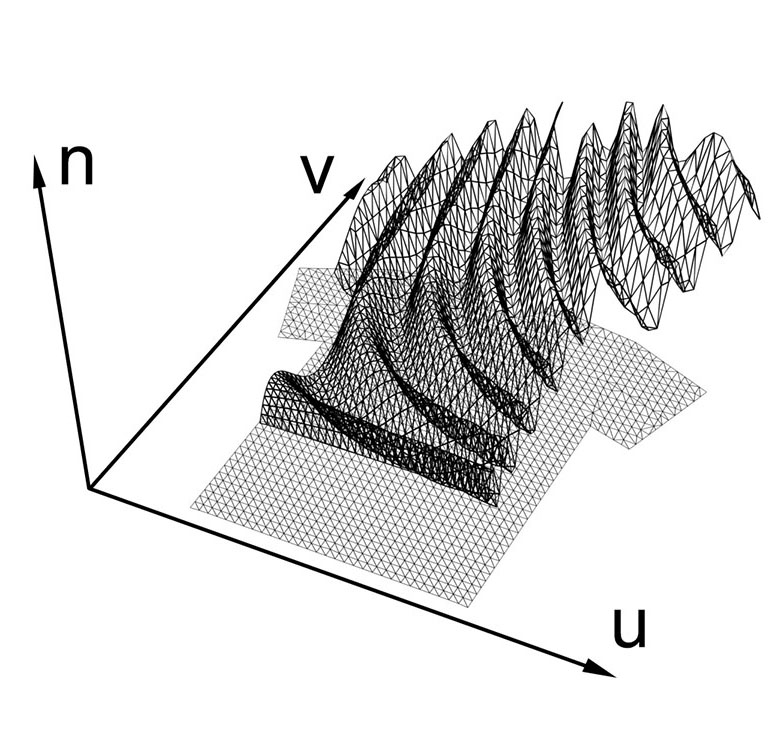}
		\caption{}
		\label{fig:front_deformed_mesh}
	\end{subfigure}
	\begin{subfigure}[b]{0.21\linewidth}
		\includegraphics[width=\linewidth]{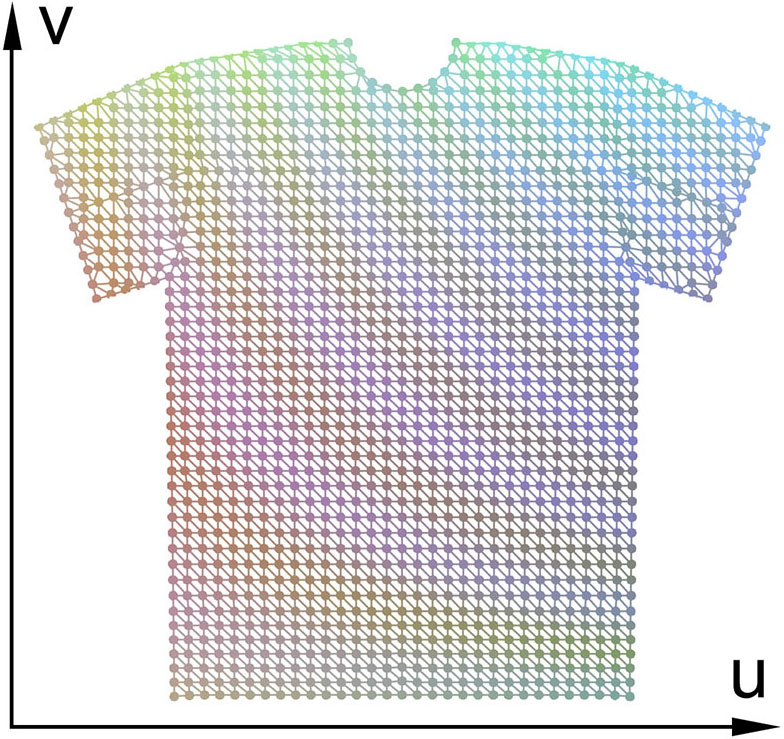}
		\caption{}
		\label{fig:front_color_deformation}
	\end{subfigure}
	\begin{subfigure}[b]{0.21\linewidth}
		\includegraphics[width=\linewidth]{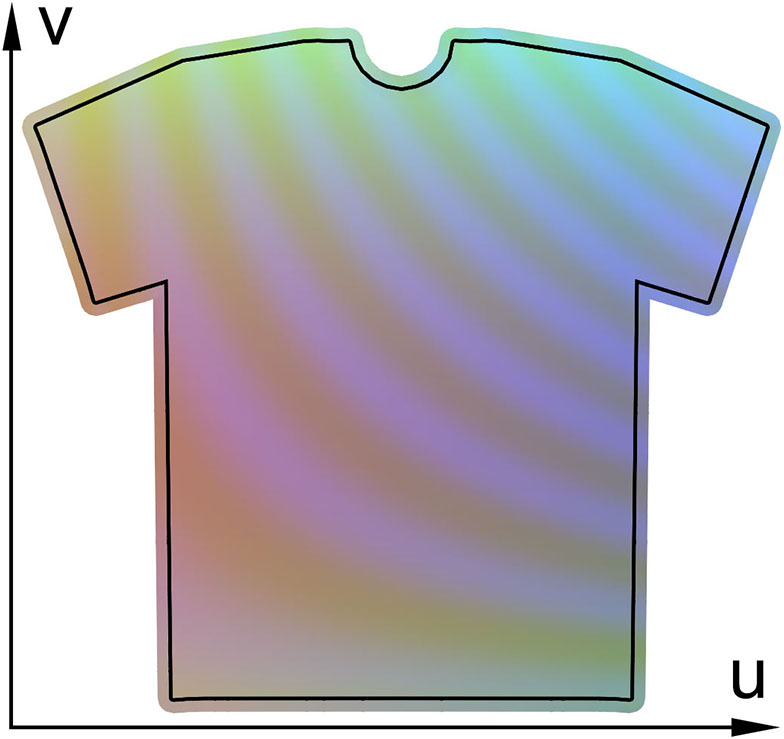}
		\caption{}
		\label{fig:front_cloth_image}
	\end{subfigure}
	\caption{(a)  A subset of the cloth mesh depicted in a two dimensional pattern space. (b) Depiction of the three dimensional displacement $d=r-\skin$ in the pattern space. (c) Displacements from (b) converted to RGB colors for each vertex. (d) Rasterization of the vertex colors from (c) onto a background grid of pixels to create an image. }
	\label{fig:deformation_image}
\end{figure}
	\renewcommand{\sim}{r}
Given $\theta$,  which may or may not be in the training set, the network needs to predict a cloth triangulated surface $\C$; however, since we keep the topology fixed, it only needs to predict vertex positions $\sim$.  
Designing an appropriate network can be quite difficult because of the non-linear joint rotations.
Thus, \cite{jin2018pixel} proposed a preprocess that utilizes a procedural skinning algorithm to obtain vertex positions that include a significant portion of the non-linear rotations. They accomplished this by shrink-wrapping the cloth vertices to the body surface $\B$ in the neutral pose (see Figure~\ref{fig:shrink_wrap_neutral}), and subsequently barycentrically embedding those cloth vertices to follow their parent triangles $b\in \B$ as the body mesh deforms (see Figure~\ref{fig:shrink_wrap_posed}). We write the simulated cloth vertex positions as $\sim=\skin+d$ where the $\skin$  one-to-one correspond to $\sim$ but are each embedded to follow a body triangle $b$, and $d$ is the remaining displacement/offset from  $\skin$ to $\sim$. 
Thus for any $\theta$, $\skin(\theta)$  is well-determined by the chosen skinning algorithm, and the network only needs to predict $d(\theta)$ from training pairs $(\theta_T,d_T)$ which  do not possess as much non-linear variation as $(\theta_T,r_T)$ do. 
Moreover, since $\skin(\theta)$ is independent of the network weights $w$, it does not need to be differentiated during numerical optimization. In summary, this can be seen as a preprocess that decomposes a function $f$ into $f_1+f_2$ where $f_1$ has less variation and is thus easier to approximate while $f_2$ has more variation but is a known function of the parameters.

Although one could train a network to interpolate from the pairs $(\theta_T,d_T)$, \cite{jin2018pixel} noted that the nature of the problem lends itself to an image based convolutional neural network (CNN) approach, and accomplished this by laying out subsets of cloth vertices in two-dimensional pattern spaces (motivated by actual garment construction from textiles). See Figure~\ref{fig:front_flat_mesh}. The three dimensional displacement $d$ can be transformed into this pattern space and displayed as a displacement of the form $(\Delta u,\Delta v,\Delta n)$ for each vertex. 
If the cloth were actually skin-tight,  $d$ would be identically zero; but otherwise, $d$ can be depicted as a new triangulated surface as shown Figure~\ref{fig:front_deformed_mesh}. Notably $(\Delta u, \Delta v, \Delta n)$ can be converted to RGB colors stored at each vertex (see Figure~\ref{fig:front_color_deformation}), and those vertex colors can be rasterized to an underlying image of pixels to obtain a cloth image $I$ (see Figure~\ref{fig:front_cloth_image}). 
Then, a convolutional neural network can be trained to interpolate from training pairs $(\theta_T,I_T)$.
Afterwards, given a pose $\theta$, the network predicts an image $I$, and the RGB colors of each cloth vertex are interpolated from this image, converted to a displacement $d$, and added to $\skin$ to obtain $\net$. 

Given training pairs $(\theta_T,I_T)$, \cite{jin2018pixel} trained their network using Adam \cite{kingma2014adam} to find network weights $w$ that minimized an energy of the form $\lVert I(\theta_T,w)-I_T\rVert$. The actual goal was to make the vertex positions match, \ie to minimize $\lVert \net(\theta_T,w)-r_T \rVert \ = \ \lVert \skin(\theta_T)+d(\theta_T,w) - (\skin_T + d_T)\rVert \ = \ \lVert d(\theta_T,w)-d_T \rVert$, and so terms of this form may be introduced as well. 
Moreover, to encourage visual similarity, \cite{jin2018pixel} also used terms that penalized differences between actual and predicted normal vectors.

%% file: section3.tex
	\section{Inextensibility Postprocess}
	\label{sec:inext_post}
\begin{figure}[b]
		\centering
		\begin{subfigure}{0.3\textwidth}
			\centering
			\includegraphics[width=0.9\textwidth]{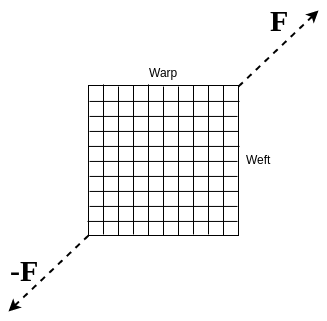}
		\end{subfigure}\hfill
		\begin{subfigure}{0.3\textwidth}
			\centering
			\includegraphics[width=0.9\textwidth]{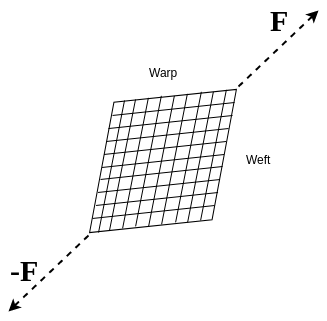}
		\end{subfigure}\hfill
		\begin{subfigure}{0.3\textwidth}
			\centering
			\includegraphics[width=0.9\textwidth]{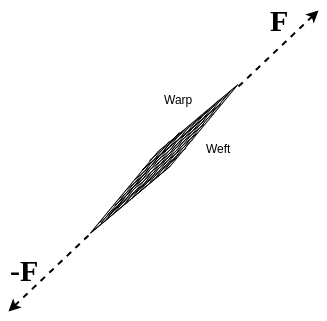}
		\end{subfigure}
		\caption{A square piece of cloth consists of generally perpendicular warp and weft threads. (a) Forces are applied to diagonal corners of a cloth patch. (b) Initially, warp and weft threads slide past each other, subject to inter-thread frictional forces. (c) As the threads align,  the resistance to deformation increases dramatically since it is difficult to stretch individual threads along their axial direction. }
		\label{fig:folding_motivation}
	\end{figure}

	Even though the network designed in \cite{jin2018pixel} predicts vertex positions as a function of joint angles quite well on average, there are a number of over-stretched/compressed elements. Thus, the cloth triangulated surface $\C$ is generally of poor physical quality and would likely behave poorly when subsequently simulated, assuming the numerical simulation would work at all. 
	As is discussed in \cite{jin2017inequality}, compression can serve as a proxy for bending/buckling that is under-resolved by the mesh;
	however, triangle edges and faces should not stretch beyond some elastic threshold regardless of the discretization.  
	This led \cite{jin2017inequality} to use different stiffnesses for compression versus extension.  
	Furthermore, they showed that strengthening per-element resistance to in-plane elastic deformation causes spurious locking (see \eg \cite{taylor1977simple}) that incorrectly removes bending degrees of freedom.
	To remedy this, they proposed a two-phase approach to elastic stretching where a reasonable strength elasticity model was used for small deformations and a constraint-based approach was used to limit larger deformations (\eg when warp and weft threads align, see Figure
	\ref{fig:folding_motivation}).
	
Following the spirit of \cite{jin2017inequality}, we replace their numerical simulation with the network prediction from \cite{jin2018pixel}, while still using the constraints proposed in \cite{jin2017inequality} to project the network output to have edge length stretching limited by a specified tolerance.
Let $\mathcal{E}$ be the set of all edges $e$ that connect two vertices in the cloth triangulated surface $\C$. Then, given a set of vertex positions $\net$ predicted by the network, they are projected to a new set of vertex positions $r$ such that each edge $\Vec{l}_e(r)$ has its length $l_e(r)\leq\lm$ where
$l_e^{\text{max}}=(1+\varepsilon)l_e^{\text{rest}}$ with $l_e^{\text{rest}}$  the rest length and $\varepsilon$ a small number. Since the solution to this problem is not unique, we make the further assumption that $r$ deviate from $\net$ as little as possible, which is justified since the network was trained to match $\net$ to the numerical simulations. In particular, we use the sum of the square of the distances between each $r$ and corresponding $\hat{r}$, which is a spring potential energy.
\begin{figure}[t]
	\centering
	\begin{subfigure}[b]{0.33\linewidth}
		\includegraphics[width=\linewidth]{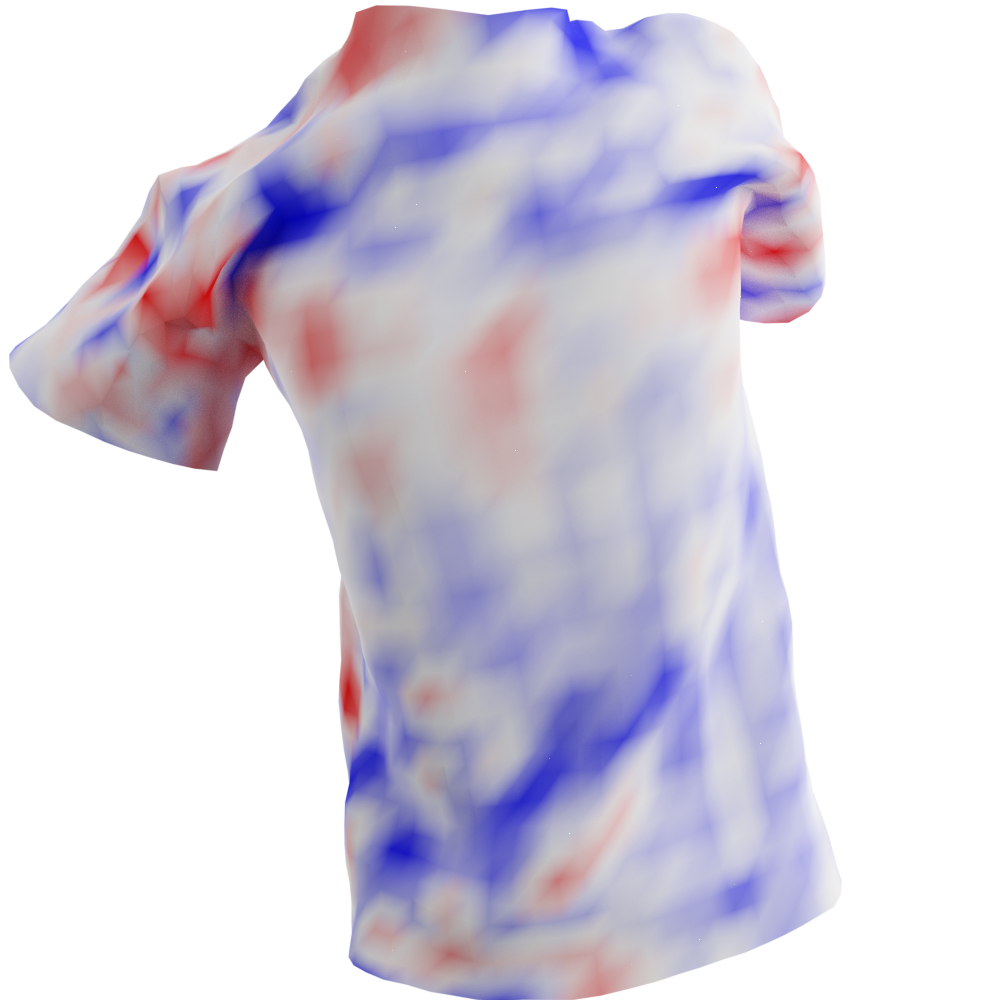}
		\caption{}
		\label{fig:pre_ineq}
	\end{subfigure}
	\begin{subfigure}[b]{0.33\linewidth}
		\includegraphics[width=\linewidth]{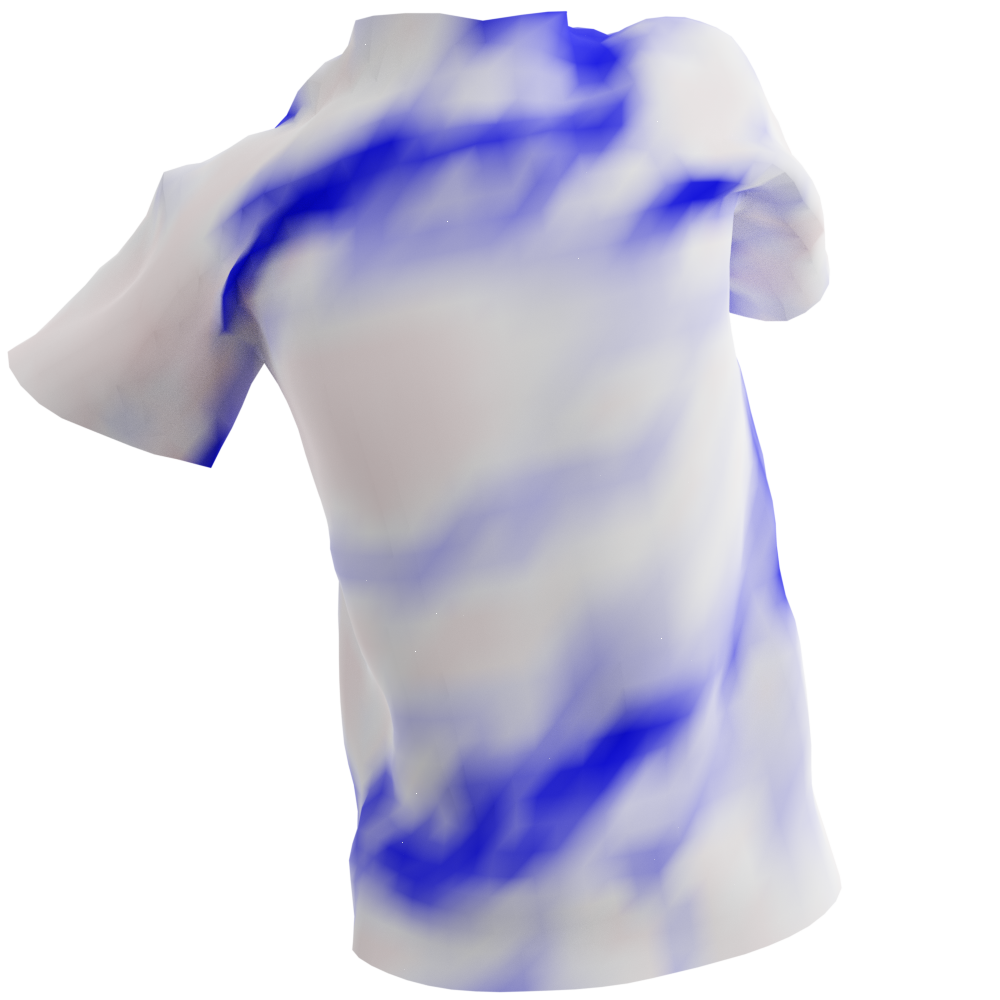}
		\caption{}
		\label{fig:post_ineq}
	\end{subfigure}
	\begin{subfigure}[b]{0.33\linewidth}
		\includegraphics[width=\linewidth]{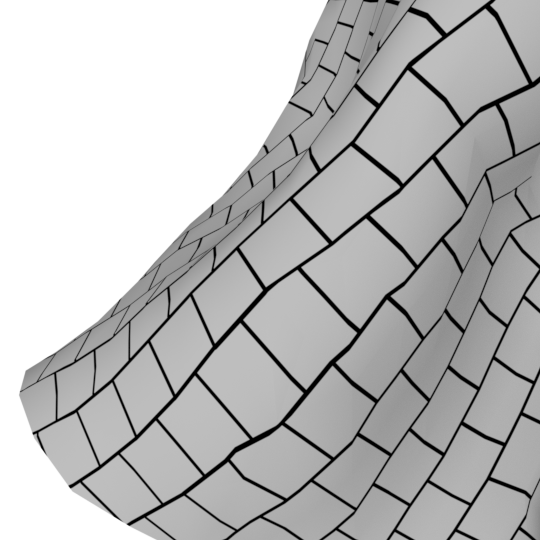}
		\caption{}
		\label{fig:pre_ineq_tex}
	\end{subfigure}
	\begin{subfigure}[b]{0.33\linewidth}
		\includegraphics[width=\linewidth]{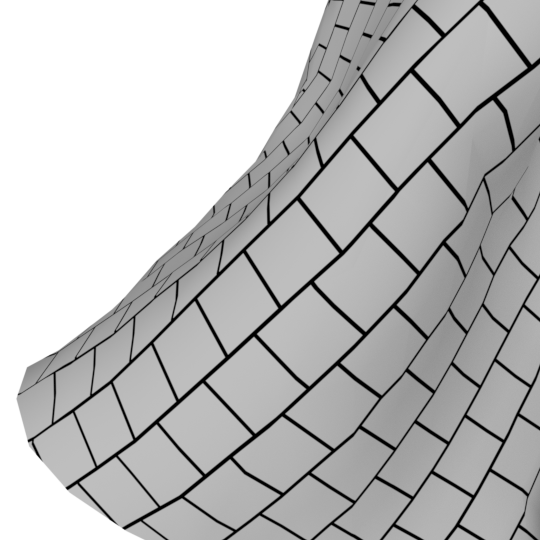}
		\caption{}
		\label{fig:post_ineq_tex1}
	\end{subfigure}
	
	\caption{(a) Network output from \cite{jin2018pixel} depicting over-stretching in red and over-compression in blue (pure red indicates 1.25 times stretching, pure blue indicates 0.75 times compression, and white indicates no distortion). (b) Results obtained after applying the inextensibility postprocess to (a). (c) Zoomed-in textured view of (a). (d) Zoomed-in textured view of (b). 
	}
	\label{fig:ineq_img}
\end{figure}	

\begin{figure}[h]
\centering
\includegraphics[width=0.45\linewidth]{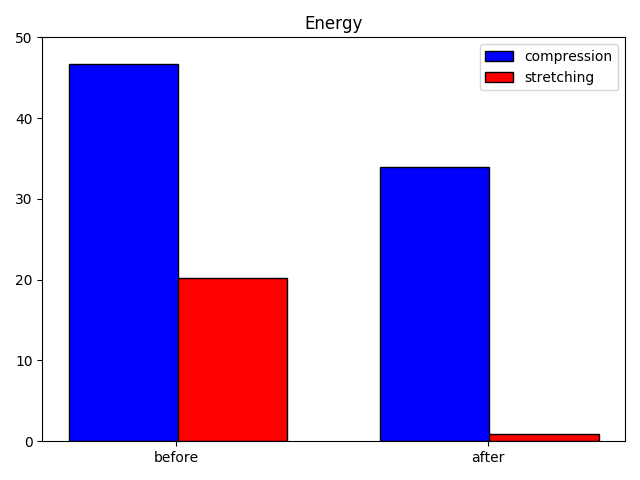}
\caption{Comparison of the energies before/after the inextensibility postprocess (averaged over all the examples in the test set).}
\label{fig:energy_ineq}
\end{figure}
	
Each vertex $r_i$ is connected to its corresponding vertex $\net_i$ via a zero-length spring with elastic force $k^z_i(\net_i-\sim_i)$ and potential energy $\frac{1}{2}k^z_i \lVert \net_i -\sim_i \rVert_2^2$ where $k^z_i$ is the spring constant. 
Since the triangulated cloth surface $\C$ is spatially adaptive with a varying mass per vertex, we set $k_i^z$ proportional to the vertex mass so that accelerations (force over mass) are more uniform.
Any overall global scaling of these forces is unimportant since it does not change the configuration where the minimum is achieved. Thus, the problem is formulated as follows:
\begin{equation}
\label{eq:inequality_min}
\begin{split}
\min_{r} &\quad \sum_i \frac{1}{2} k_i^z \lVert \net_i - \sim_i \rVert_2^2 \\
\text{subject to}  &\quad l_e(r)\leq \lm, \quad \forall e \in \mathcal{E}.
\end{split}
\end{equation}
This is a convex minimization problem with a unique solution, and has KKT conditions \cite{boyd2004convex}:
\begin{equation}
\label{eq:inequality_cloth_KKT}
\begin{gathered}
\lambda_e \geq 0 ,\quad l_e(r)-\lm \leq 0,\quad \lambda_e(l_e(r)-\lm)=0, \quad \forall e \in \E \\
k_i^z(\net_i-\sim_i) +\sum_{e\in \E_i} \lambda_e \hat{l}_e^{\ i}(r)=0, \quad \forall i
\end{gathered}
\end{equation}
where $\lambda_e$ are Lagrange multipliers which may only be non-zero when $l_e(r)=\lm$. $\E_i$ consists of all the edges from $\E$ that include vertex $i$, reoriented (if necessary) so that vertex $i$ appears first, and $\hat{l}_e^{\ i}(r)$ is the unit vector that points from $r_i$ to the location of the other vertex on the edge $e\in \E_i$. When non-zero, the equal and opposite per-edge constraint forces act as strong edge springs that balance the zero length springs into quasi-static equilibrium while enforcing the constraints.   

\subsection{Examples}
\label{sec:inext_post_examples}
	Equation~\ref{eq:inequality_min} can be recast into a second-order cone program (see Section~\ref{sec:inext_prior}) and solved efficiently with the method from \cite{domahidi2013ecos}. Figure~\ref{fig:pre_ineq} shows a typical result output from the network proposed in \cite{jin2018pixel}, depicting over-stretched edges in red and over-compressed edges in blue. Figure~\ref{fig:post_ineq} shows the result obtained by solving Equation~\ref{eq:inequality_min} as a postprocess to the network result shown in Figure~\ref{fig:pre_ineq}. Figure~\ref{fig:pre_ineq_tex} shows a zoomed-in textured view of the right sleeve in Figure~\ref{fig:pre_ineq}. Figure~\ref{fig:post_ineq_tex1} shows a zoomed-in textured view of the same area of Figure~\ref{fig:post_ineq}, highlighting how the postprocess alleviates in-plane distortions.
	Figure~\ref{fig:energy_ineq} shows a comparison of the spring energies before and after the inextensibility postprocess (averaged over all examples in the test set) illustrating a drastic removal of over-stretching. 

%% file: section4.tex
\section{Inextensibility Prior}
	\label{sec:inext_prior}
	
	Assuming $\opt$ is the unique solution to Equation~\ref{eq:inequality_min}, embedding the postprocess of Section~\ref{sec:inext_post} into the neural network is accomplished by
	changing the energy to be minimized replacing terms of the form $\lVert \net-r_T\rVert_2^2$ with $\lVert \opt-r_T \rVert_2^2$. Then, the minimization process requires the derivatives of this new energy with respect to the network weights $w$.
	Note that $\frac{\partial \net}{\partial w}$ is readily accessible since it was used to train the network in \cite{jin2018pixel}. In addition, the derivatives of the energy with respect to $\opt$ are typically straight-forward. Thus, we only need discuss $\frac{\partial \opt}{\partial \net}$;
	however,$\frac{\partial \opt}{\partial \net}$ is not altogether clear from Equations~\ref{eq:inequality_min} and \ref{eq:inequality_cloth_KKT}. 
	In order to better elucidate $\frac{\partial \opt}{\partial \net}$, we consider the solution process proposed in \cite{domahidi2013ecos}. 
	
	Equation~\ref{eq:inequality_min} is rewritten in conic form as follows.
	Let $Q^m$ be the space of second order cones with dimension $m$, \ie  $[q_0,q_1^T]^T\in Q^{m}$ if and only if $\lVert q_1 \rVert_2\leq q_0$ where $q_0\in \mathbb{R},\ q_1\in \mathbb{R}^{m-1}$. Then defining $s_0=[t,\sqrt{k_i^z}(\sim_i-\net_i)^T,\dots]^T $ allows the objective function in Equation~\ref{eq:inequality_min} to be recast as minimizing the slack variable $t$ while maintaining $s_0\in Q^{3n+1}$. Similarly, defining $s_e=[\lm,\Vec{l}_e(r)^T]^T$ allows the constraints to be represented by simply stating that $s_e\in Q^4$ for all $e \in \E$. Alternatively, one can specify $s_e=[\alpha_e,\Vec{l}_e(r)^T]^T$ where the $\alpha_e$ are additional variables constrained via $\alpha_e=\lm$. Concatenating all the $\alpha_e$ into a single vector $\alpha$, we obtain
	\begin{equation}
	\label{eq:inequality_convert}
	\begin{split}
	\min_{r,t,\alpha} & \quad t \\
	\text{subject to} & \quad s_0 \in Q^{3n+1} \\
	&\quad s_e \in Q^{4}, \quad \forall e \in \E \\
	&\quad \alpha_e=\lm,  \quad \forall e \in \E.
	\end{split}
	\end{equation}
	To simplify, let $x=[r^T,t,\alpha^T]^T$. Similarly, concatenate $s_0$ and all the $s_e$ into a single vector $s$, 
	defining $\K$ as the space of all $s$ that have sub-vector $s_0\in Q^{3n+1}$ and all sub-vectors $s_e\in Q^4$.
	With this notation, Equation~\ref{eq:inequality_convert} can be written more formally as a second order cone program (SOCP):
	\begin{equation}
	\label{eq:SOCP_primal}
	\begin{split}
	\min_{x,s} \quad & c^Tx \\
	\text{subject to} \quad & Ax=b \\
	\quad& s\in \K \\
	\quad& s=h-Gx.
	\end{split}
	\end{equation}
	Here, $c$ merely selects $t$ from $x$, and thus does not depend on $\net$. $A$ selects all $\alpha_e$ from $x$ and sets them equal to the corresponding $\lm$ in $b$, and so neither $A$ nor $b$ depends on $\net$. The constraints $s_0\in Q^{3n+1}$ and $s_e\in Q^4$ for $\forall e \in \E$ are folded into $s\in\K$. Finally, $s=h-Gx$ ties $s$ and $x$ together, defining $s_0$ and all $s_e$. Although $s_e$ is defined by $G$ alone, $s_0$ is defined by both $G$ and $h$ with $-\sqrt{k_i^z}\net_i$ terms in $h$ making $h$ the only term in Equation~\ref{eq:SOCP_primal} that depends on $\net$. 
	Thus, we may write $\frac{\partial \opt}{\partial \net}$ as $\frac{\partial \opt}{\partial h}\frac{\partial h}{\partial \net}$, where $\frac{\partial h}{\partial \net}$ has non-zero terms of the form $-\sqrt{k_z^i}$.
	Equation~\ref{eq:SOCP_primal} is the primal form of the second order cone program, and the dual form is 
	\begin{equation}
	\label{eq:SOCP_dual}
	\begin{split}
	\max_{y,z} \quad & -b^Ty-h^Tz \\
	\text{subject to} \quad & G^Tz+A^Ty+c=0 \\
	\quad &  z\in \mathcal{K}.
	\end{split}
	\end{equation}
	As discussed in \cite{domahidi2013ecos}, the primal and dual problems are optimal at the same point. All the constraints from Equations~\ref{eq:SOCP_primal} and \ref{eq:SOCP_dual} can be collected to write
	\begin{equation}
	\label{eq:SOCP_KKT}
	\begin{bmatrix}
	0 & A^T & G^T & 0\\
	A & 0 & 0 & 0\\
	G & 0 & 0 & I\\
	\end{bmatrix}
	\begin{bmatrix}
	x \\ y \\ z \\ s
	\end{bmatrix}
	=
	\begin{bmatrix}
	-c\\
	b\\
	h
	\end{bmatrix},
	\quad s,z \in \K.
	\end{equation}
	Next, given $q=[q_0,q_1^T]^T$ and $\hat{q}=[\hat{q}_0,\hat{q}_1^T]^T$, the conic product is defined as $q\circ \hat{q}=\begin{bmatrix}
	q^T\hat{q}\\
	q_0\hat{q}_1+\hat{q}_0q_1
	\end{bmatrix}$.
	For the concatenated cone space $\K$, the conic product $s\circ z$ is defined via separate conic products between each pair of component cones.
	As discussed in \cite{domahidi2013ecos}, the joint solution to the primal and dual problems satisfies the constraints in Equation~\ref{eq:SOCP_KKT} along with $s\circ z=0$.
	
	Taking differentials of the linear system in Equation~\ref{eq:SOCP_KKT} as well as $s\circ z=0$ yields
	\begin{equation}
	\label{eq:inext_differential}
	\begin{bmatrix}
	0 & A^T & G^T & 0 \\
	A & 0 & 0 & 0\\
	G & 0 & 0 & I \\
	0 & 0 & S & Z
	\end{bmatrix}
	\begin{bmatrix}
	dx \\ dy \\ dz \\ds
	\end{bmatrix}
	=
	\begin{bmatrix}
	0\\ 0\\ dh \\ 0
	\end{bmatrix}
	\end{equation}
	where $S$ and $Z$ are the matrices defined via $S dz=s\circ dz$ and $Z ds=z\circ ds\ (=ds\circ z)$ respectively, so that the last equation is 
	$S dz+ Z ds=0$. 
	
The interior point solution method of \cite{domahidi2013ecos} provides a point which is nearly optimal, but not exactly so due to numerical errors, the stopping conditions/tolerance on iterations, and the need to project the solution/iterates into the domain interior. Thus, their solution will approximately satisfy the optimality conditions, and the differentials evaluated at their solution will approximately satisfy Equation~\ref{eq:inext_differential}. Although Equation~\ref{eq:inext_differential} is generally not full rank, we discuss its null space in Appendix A so that one may still find suitable values for the differentials in a minimal norm sense. Notably, the interior point method of \cite{domahidi2013ecos} returns a solution where the coefficient matrix in Equation~\ref{eq:inext_differential} is full rank.
	
To symmetrize Equation~\ref{eq:inext_differential}, we column scale defining a new variable $d\psi$ such that $ds=Sd\psi$ to obtain
	\begin{equation}
	\label{eq:sym_system}
	\begin{bmatrix}
	0 & A^T & G^T & 0 \\
	A & 0 & 0 & 0 \\
	G & 0 & 0 & \So \\
	0 & 0 & \So & \Zo\So
	\end{bmatrix}
	\begin{bmatrix}
	d\xo \\ d\yo \\ d\z \\ d\psi
	\end{bmatrix}
	=
	\begin{bmatrix}
	0\\ 0\\ dh \\ 0
	\end{bmatrix}.
	\end{equation}
	Both $\Zo,\So$ are symmetric, and their product $\Zo\So$ \textit{should} be symmetric as well. Since $Z,S$ and $ZS$ are all block diagonal, each block can be considered independently. Let $q$ and $\hat{q}$ be sub-cones, and define $M_q$ so that $M_q\hat{q}=q\circ\hat{q}$, \ie
	\begin{equation}
	\label{eq:sec3:conic_product}
	M_q=\begin{bmatrix}
	q_0 & q_1^T \\
	q_1 & q_0 I
	\end{bmatrix}.
	\end{equation}
	Then, a diagonal block of $\Zo\So$ has the form:
	\begin{equation}
	M_qM_{\hat{q}}=\begin{bmatrix}
	q_0\hat{q}_0+q_1^T\hat{q}_1 & q_0\hat{q}_1^T+\hat{q}_0q_1^T \\
	q_0\hat{q}_1+\hat{q}_0q_1 & q_1\hat{q}_1^T+q_0\hat{q}_0 I
	\end{bmatrix}
	\end{equation}
	where the upper left hand corner is a scalar, and the off diagonal terms $q_0\hat{q}_1+\hat{q}_0q_1$ and its transpose should both be identically $0$ under the optimality constraints. The lower right hand term has a symmetric component $q_0\hat{q}_0I$ and an outer product component $q_1\hat{q}_1^T$. If $q_0$ or $\hat{q}_0$ is $0$, then all of $q_1$ or $\hat{q}_1$ respectively is also zero, making the entire term on the lower right hand block of $M_qM_{\hat{q}}$ equal to zero, so the block is trivially symmetric. In practice, the interior solver \cite{domahidi2013ecos} returns a point on the interior of $\K$, so both $q_0$ and $\hat{q}_0$ are non-zero. Using the optimality condition $q_0\hat{q}_1+\hat{q}_0q_1=0$, one can rewrite $q_1\hat{q}_1^T$ as either $-\frac{\hat{q}_0}{q_0}q_1q_1^T$ or $-\frac{q_0}{\hat{q}_0}\hat{q}_1\hat{q}_1^T$ depending on whether $q_0$ or $\hat{q}_0$ is larger (for robustness). In summary, we symmetrize each diagonal block in this manner, projecting away  numerical errors from the interior solver of \cite{domahidi2013ecos} that lead to small asymmetries. 
	
	Denoting the symmetrized version of the coefficient matrix in Equation~\ref{eq:sym_system} as $K$ and the unknown vector as $d\xi$, Equation~\ref{eq:sym_system} should be solved separately for each variable in $h$. This is accomplished by dividing both sides by $dh_j$ to obtain unknown $\frac{\partial \xi}{\partial h_j}$ with basis vector $[0,0,e_j^T,0]^T$ on the right hand side. This can be conveniently notationally stacked into the expression $K\frac{\partial \xi}{\partial h}=I_h$ where $\frac{\partial \xi}{\partial h}$ consists of columns of the form $\frac{\partial \xi}{\partial h_j}$ and $I_h$ is a block column matrix of zeros except for an identity matrix corresponding to the location of $h$. Since $\opt$ is a sub-component of $x^*$, we may write $\frac{\partial \opt}{\partial h}=I_r \frac{\partial \xi}{\partial h}$ where $I_r$ is a block row matrix of zeros with an identity matrix corresponding to the location of $\opt$. 
	In summary,
	\begin{equation}
	\label{eq:sec4_dE_dw}
	\pd{\Et}{w}= \pd{E}{\opt}\pd{\opt}{\xi}\pd{\xi}{h}\pd{h}{\net}\pd{\net}{w}=\pd{\Et}{\opt}I_rK^{-1}I_h\pd{h}{\net}\pd{\net}{w}=\left(\pd{\Et}{\opt}I_rK^{-1}\right)I_h\pd{h}{\net}\pd{\net}{w}
	\end{equation}
	where $\pd{\Et}{\opt}$ is straight-forward to calculate based on the form of the energy function, $\pd{h}{\net}$ is based on the definition of $h$ in Equation~\ref{eq:SOCP_primal}, and $\pd{\net}{w}$ is available from the network in \cite{jin2018pixel}.
	For efficiency, note that $\pd{\Et}{\opt}$ and $\pd{\Et}{\opt} I_r$ are row vectors, and so we may compute the row vector $\eta=\pd{\Et}{\opt} I_r K^{-1}$ by solving $K\eta^T=I_r^T \left(\pd{\Et}{\opt}\right)^T$ once for each training example; in fact, we use sparse LDL similar to \cite{domahidi2013ecos}.

\subsection{Examples}
\label{sec:inext_prior_examples}
We use the cloth data set from \cite{jin2018pixel} to obtain both the training examples and ground truth consistent with our approach. The cloth consists of 2969 vertices with 8787 inequality constraints. We minimize an energy of the form $\lVert \opt-r_T \rVert$, whereas \cite{jin2018pixel} minimized an energy of the form $\lVert \hat{I}-I_T \rVert$; thus, we retrain the same network from \cite{jin2018pixel} using $\lVert \net-r_T\rVert$. The derivatives of the total energy with respect to the weights $w$ can be computed separately for the terms corresponding to each training example using Equation~\ref{eq:sec4_dE_dw}. For each training example, solving Equation~\ref{eq:inequality_min} takes 3-4 seconds and computing $\frac{\partial E}{\partial \net}$ via $\eta$ takes about $0.5$ seconds.  


\begin{table}[h]
\centering
\begin{tabular}{c|c|c|c|c|c|c}
\hline
\multirow{2}{*}{Method} & \multicolumn{2}{c|}{Training Set} & \multicolumn{2}{c|}{Validation Set} & \multicolumn{2}{c}{Test Set} \\ \cline{2-7}
& SqrtMSE  & MaxDist  & SqrtMSE & MaxDist & SqrtMSE & MaxDist  \\  \hline 
\cite{jin2018pixel} & $2.5370$ & $10.554$ & $7.6060$ & $29.799$ & $7.6840$ & $29.637$ \\ \hline
Postprocess Only & $2.0781$ & $10.042$ & $7.2026$ & $28.880$ & $7.4326$ & $29.012$ \\ \hline
Trained Postprocess & $2.0767$ & $9.9807$ & $7.1903$ & $29.181$ & $7.1603$ & $29.006$ \\ \hline
\end{tabular}
\caption{Errors on a data set with $1000$ training examples in the energy function. Numbers are in millimeters.}
\label{tab:sec4_1k}
\end{table}

\begin{table}[h]
\centering
\begin{tabular}{c|c|c|c|c|c|c}
\hline
\multirow{3}{*}{Method} & \multicolumn{2}{c|}{Training Set} & \multicolumn{2}{c|}{Validation Set} & \multicolumn{2}{c}{Test Set} \\ \cline{2-7}
& SqrtMSE  & MaxDist  & SqrtMSE & MaxDist & SqrtMSE & MaxDist  \\  \hline 
\cite{jin2018pixel} & $4.1824$ & $17.351$ & $5.9018$ & $25.506$ & $5.9405$ & $25.597$ \\ \hline
Postprocess Only & $3.9257$ & $16.853$ & $5.6153$ & $25.502$ & $5.6582$ & $25.215$ \\ \hline
Trained Postprocess & $3.8096$ & $16.790$ & $5.5875$ & $25.030$ & $5.6511$ & $25.267$ \\ \hline
\end{tabular}
\caption{Errors on a data set with $8000$ training examples in the energy function. Numbers are in millimeters.}
\label{tab:sec4_8k}
\end{table}
We train our network on a data set with $1000$ training examples and report the errors in Table~\ref{tab:sec4_1k}. We use two error metrics: square root of mean square error (SqrtMSE) and maximum vertex distance (MaxDist). We compare errors for three approaches: results obtained using the network from \cite{jin2018pixel} (but with $\lVert \net-r_T\rVert$), applying the inextensibility postprocess only to the results of \cite{jin2018pixel}, and training with the inextensibility postprocess in the network. The training set is the set of examples used in the energy function. At training time, we periodically evaluate the network on a validation set and save the network weights $w$ that have the smallest SqrtMSE. The test set of data is not seen during training and is used as a proxy for measuring generalization error.  As can be seen in the table, the inextensibility postprocess reduces the errors in all cases, ranging from small improvements up to around $25\%$; however, including the postprocess in training generally yields only minor improvements. The results obtained by incorporating the postprocess in training are similar enough to those obtained using the postprocess only that the comments in Section~\ref{sec:inext_post_examples} and Figures~\ref{fig:ineq_img} and \ref{fig:energy_ineq} are representative of both. Table~\ref{tab:sec4_8k} shows results one would expect when increasing the number of training examples from $1000$ to $8000$: it is harder to reduce the energy, so the training set errors increase; however, having more training examples reduces the errors on the validation set and test set.

%% file: section5.tex
\section{Buckling and Inextensibility Postprocess}
\label{sec:buck_post}
Although the inextensibility postprocess discussed in Sections~\ref{sec:inext_post} and \ref{sec:inext_prior} adequately prevents overstretching, it still allows for compression as motivated by \cite{jin2017inequality}; however, there are often times when it is desirable to limit element compression as well. Thus, we introduce a quasistatic simulation postprocess to better capture buckling phenomena in addition to inextensibility. Although there are a variety of material models one might employ,  the numerical method should be robust enough to handle the poorly posed initial conditions output by the network including poorly conditioned and inverted elements/configurations.

\begin{figure}[h]
	\centering
	\includegraphics[width=0.5\textwidth]{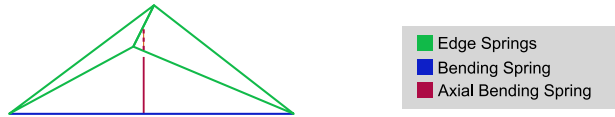}
	\caption{A pair of triangles in the cloth mesh (green). Besides using springs on the edges of the original cloth mesh, bending springs (blue) are created for fictitious edges connecting the non-shared vertices for every pair of triangles. In order to prevent ill-conditioned oscillations near a flat rest state, a zero-length spring (red) connects the fictitious bending edge with the edge shared by the pair of triangles. See \cite{selle2008mass}. }
	\label{fig:sec5_springs}
\end{figure}

\begin{figure}[h!]
	\centering
	\begin{minipage}{\textwidth}
		\begin{subfigure}[b]{\stiffnessfigwidth\linewidth}
			\includegraphics[width=\linewidth]{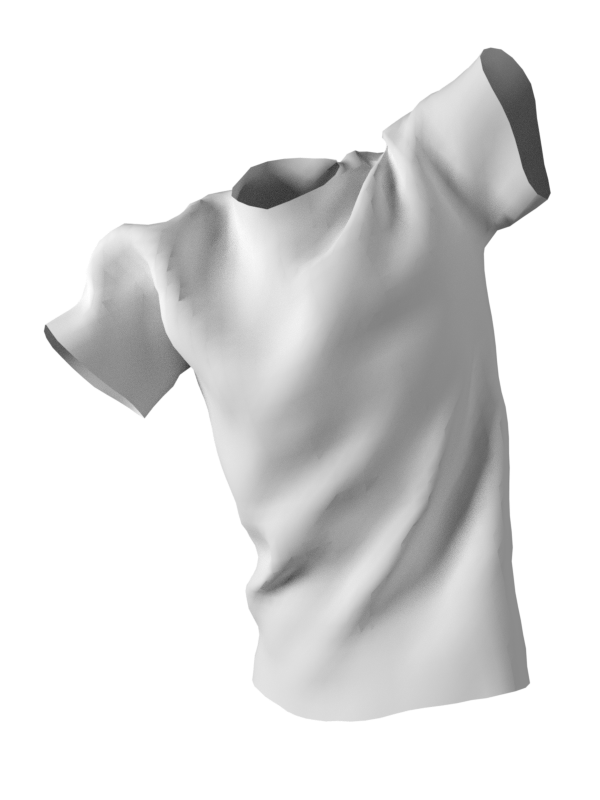}
			\caption{output from \cite{jin2018pixel}}
			\label{fig:stiffness_pd}
		\end{subfigure}
		\hspace{1mm}
		\begin{subfigure}[b]{\stiffnessfigwidth\linewidth}
			\includegraphics[width=\linewidth]{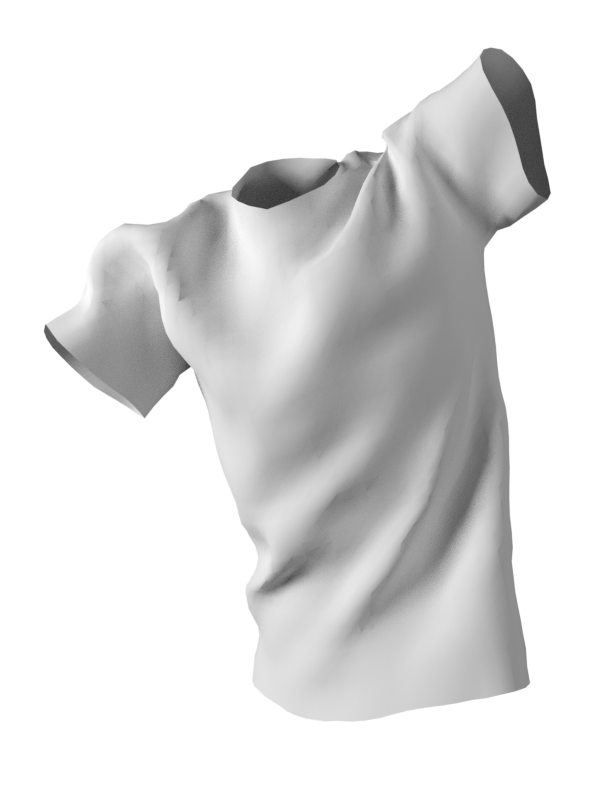}
			\caption{$\mu=10^3$}
			\label{fig:stiffness_big}
		\end{subfigure}
		\begin{subfigure}[b]{\stiffnessfigwidth\linewidth}
			\includegraphics[width=\linewidth]{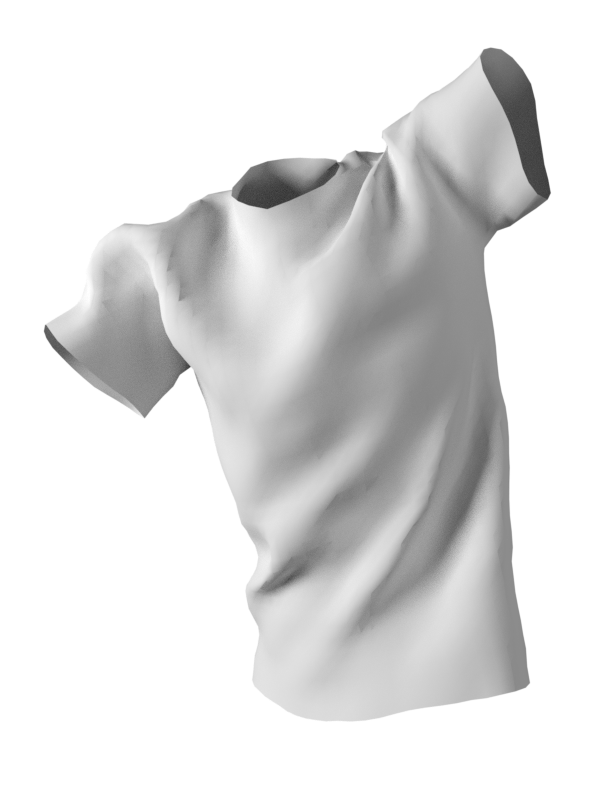}
			\caption{$\mu=10^2$}
			\label{}
		\end{subfigure}
		\begin{subfigure}[b]{\stiffnessfigwidth\linewidth}
			\includegraphics[width=\linewidth]{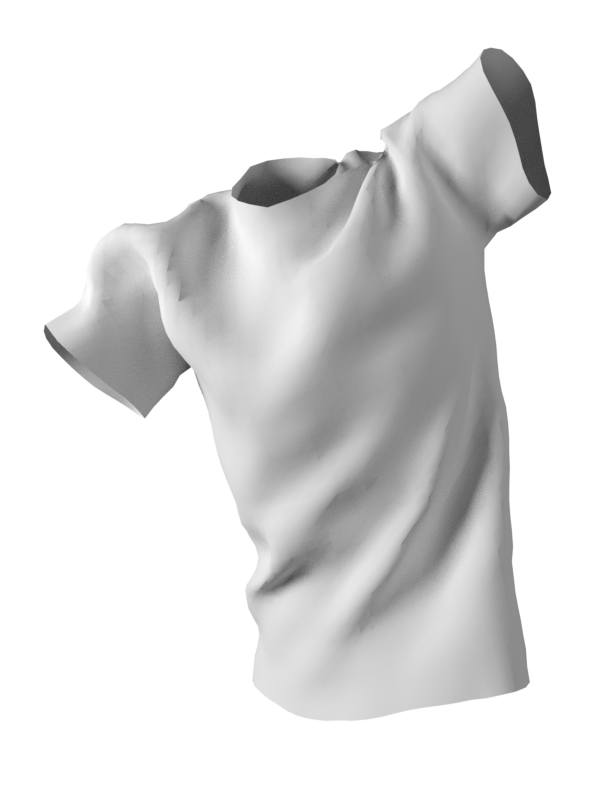}
			\caption{$\mu=10^1$}
			\label{}
		\end{subfigure}
		\begin{subfigure}[b]{\stiffnessfigwidth\linewidth}
			\includegraphics[width=\linewidth]{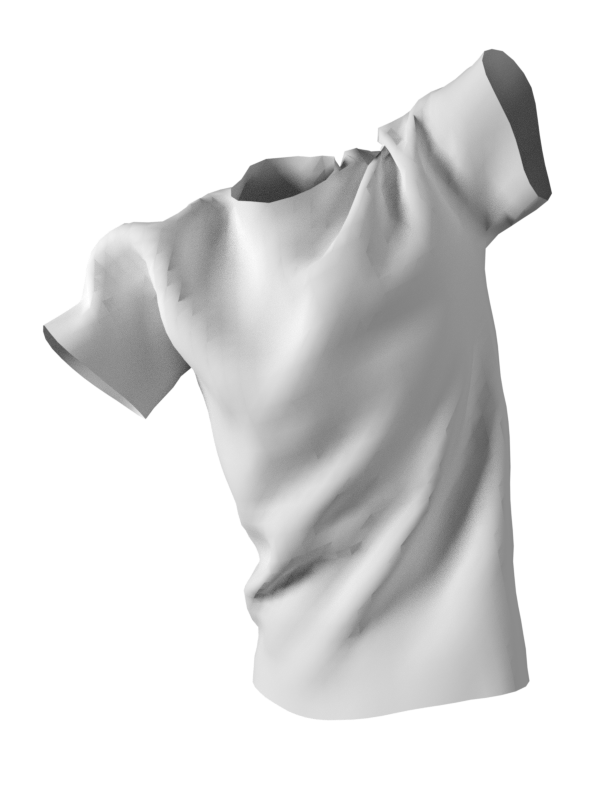}
			\caption{$\mu=10^{0}$}
			\label{}
		\end{subfigure}
		\begin{subfigure}[b]{\stiffnessfigwidth\linewidth}
			\includegraphics[width=\linewidth]{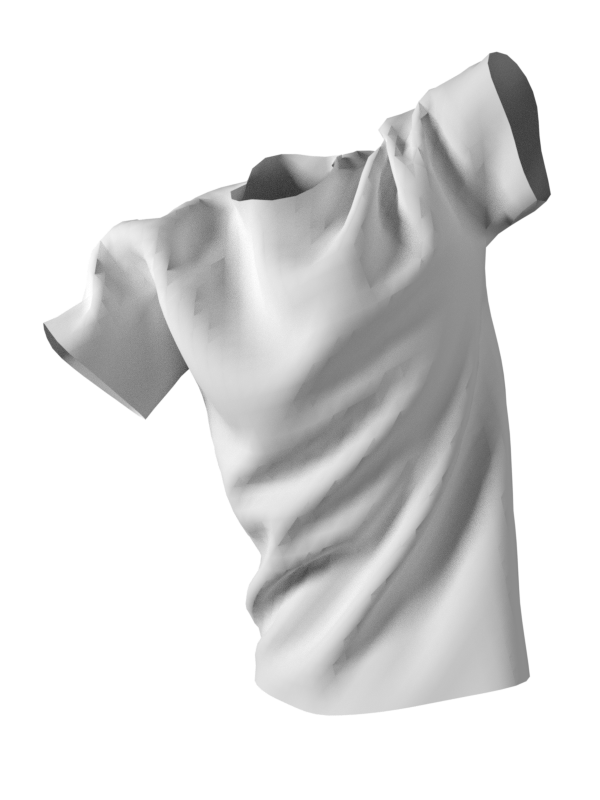}
			\caption{ground truth}
			\label{fig:stiffnes_gt}
		\end{subfigure}
		\hspace{1mm}
		\begin{subfigure}[b]{\stiffnessfigwidth\linewidth}
			\includegraphics[width=\linewidth]{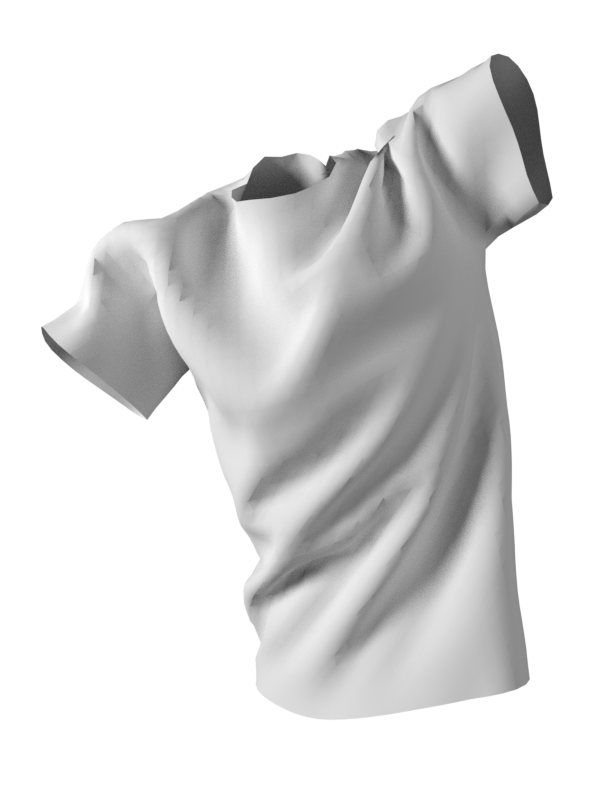}
			\caption{$\mu=10^{-1}$}
			\label{}
		\end{subfigure}
		\begin{subfigure}[b]{\stiffnessfigwidth\linewidth}
			\includegraphics[width=\linewidth]{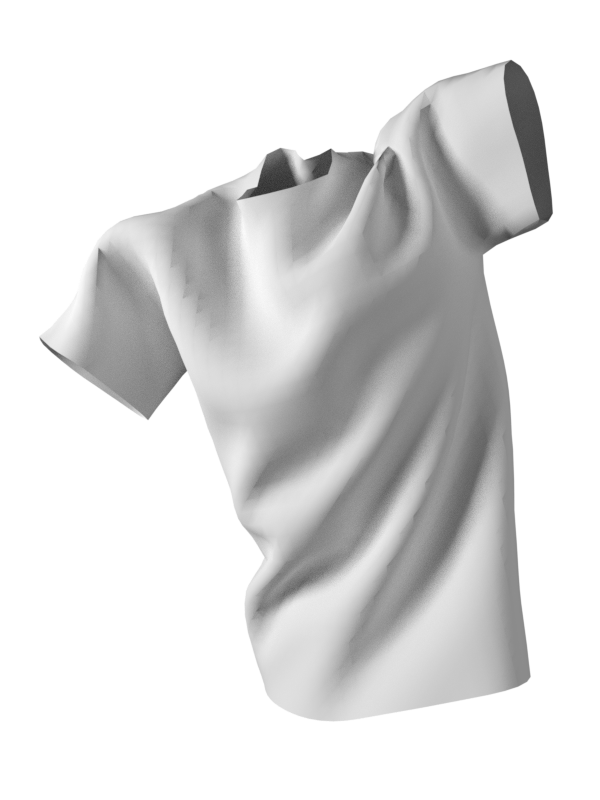}
			\caption{$\mu=10^{-2}$}
			\label{}
		\end{subfigure}
		\begin{subfigure}[b]{\stiffnessfigwidth\linewidth}
			\includegraphics[width=\linewidth]{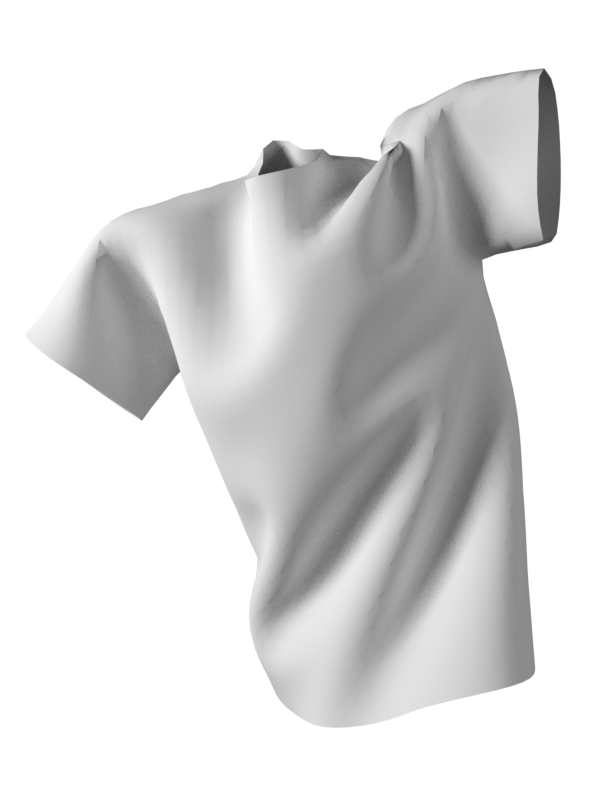}
			\caption{$\mu=10^{-3}$}
			\label{}
		\end{subfigure}
		\begin{subfigure}[b]{\stiffnessfigwidth\linewidth}
			\includegraphics[width=\linewidth]{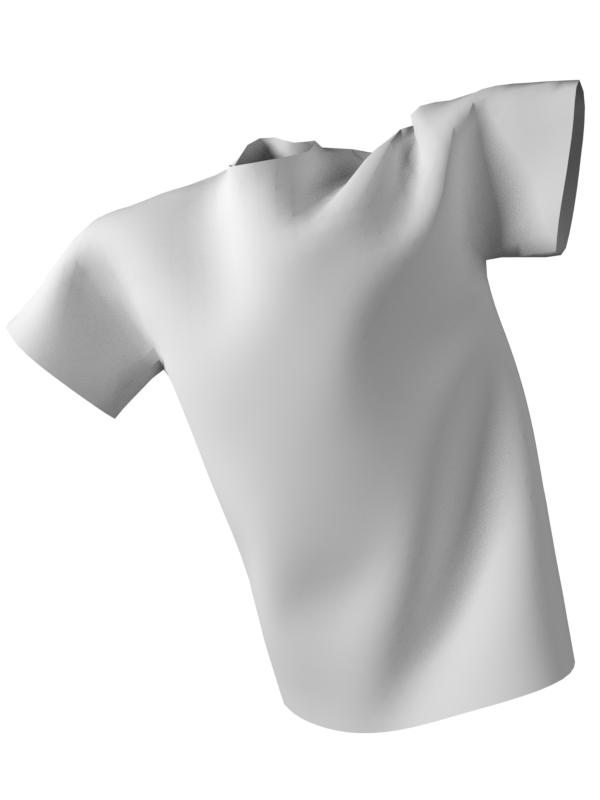}
			\caption{$\mu=10^{-4}$}
			\label{fig:stiffness_small}
		\end{subfigure}
	\end{minipage}
	\caption{Comparison of the ground truth (f), the network result (a),  and a number of quasistatic postprocessed solutions. Note how the stiffest zero length springs in (b) produce a result close to \cite{jin2018pixel}, and how the weaker zero-length springs allow the cloth to drift too far from the ground truth. Also note how well the quasistatic solution shown in figure (g) matches the ground truth shown in (f).  }
	\label{fig:stiffness}
\end{figure}

\begin{figure}[h!]
	\centering
	\begin{subfigure}[b]{0.32\linewidth}
		\includegraphics[width=\linewidth]{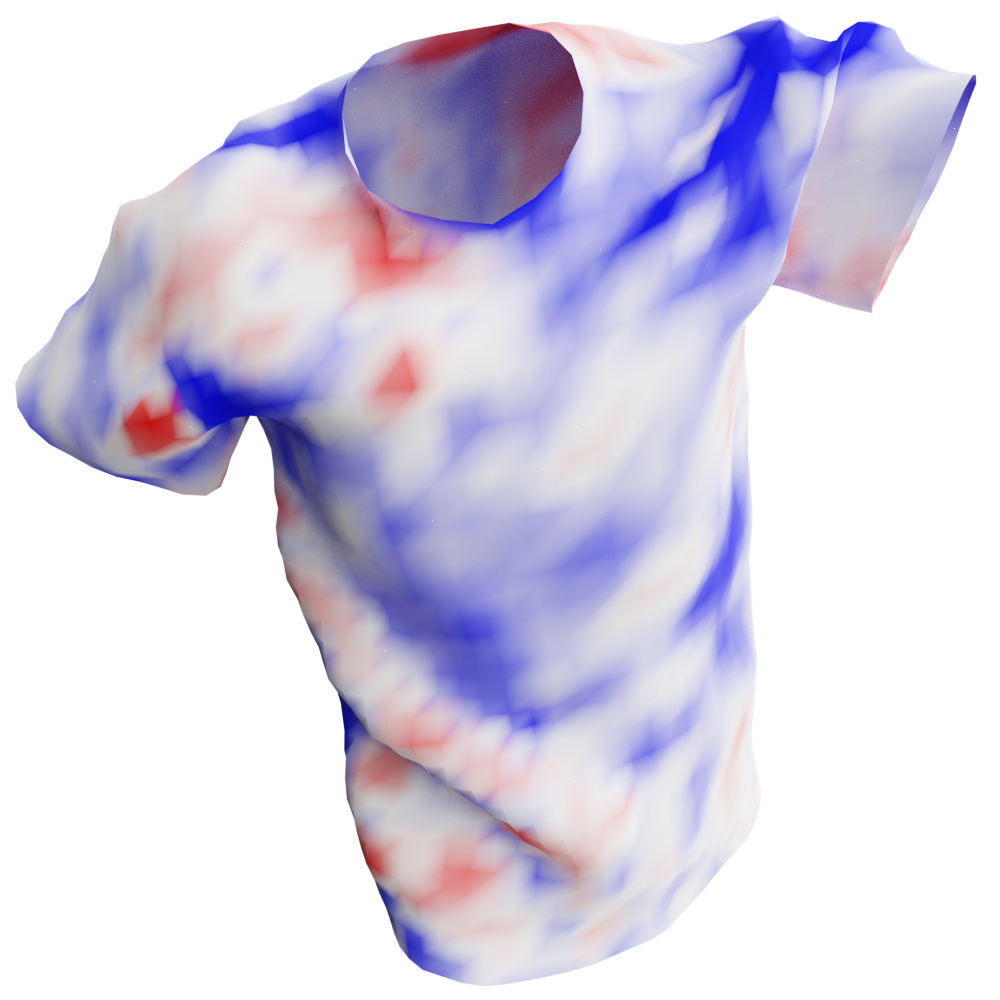}
		\caption{}
		\label{fig:pre_buck}
	\end{subfigure}
	\begin{subfigure}[b]{0.32\linewidth}
		\includegraphics[width=\linewidth]{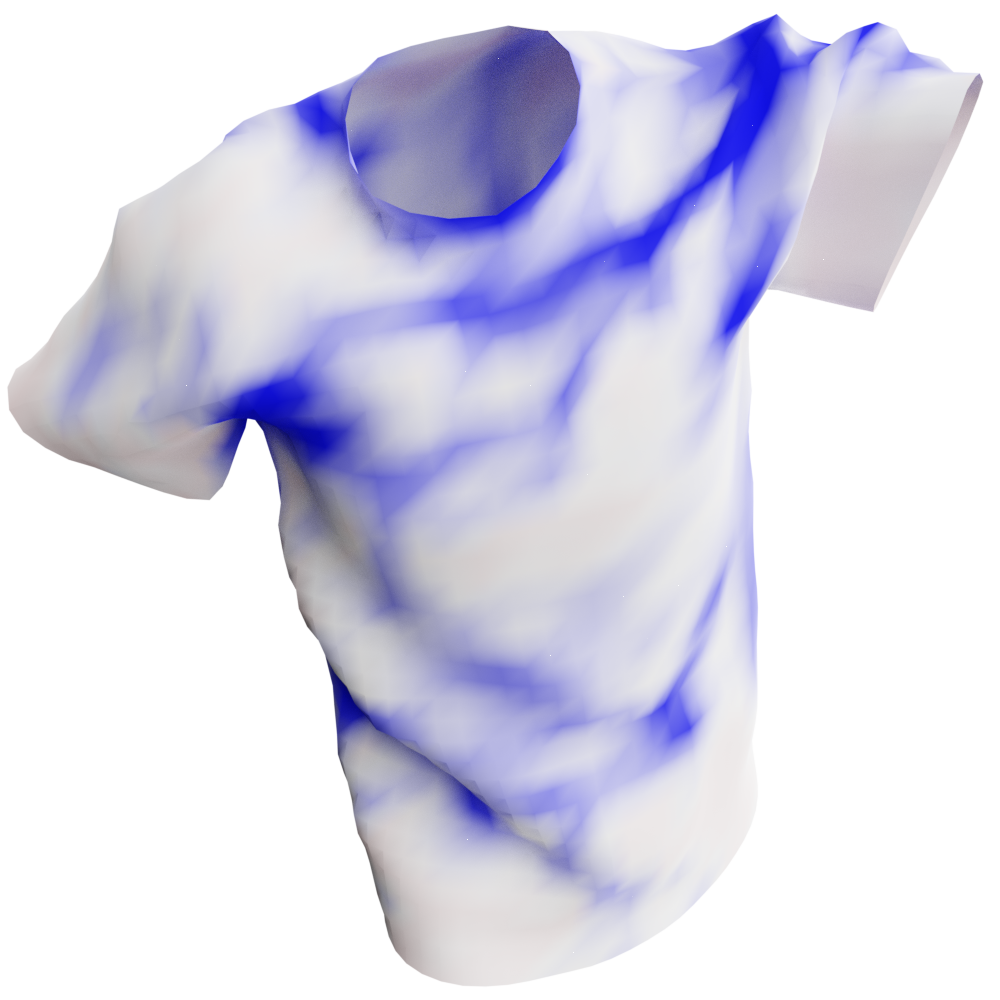}
		\caption{}
		\label{fig:post_ineq_buck}
	\end{subfigure}
	\begin{subfigure}[b]{0.32\linewidth}
		\includegraphics[width=\linewidth]{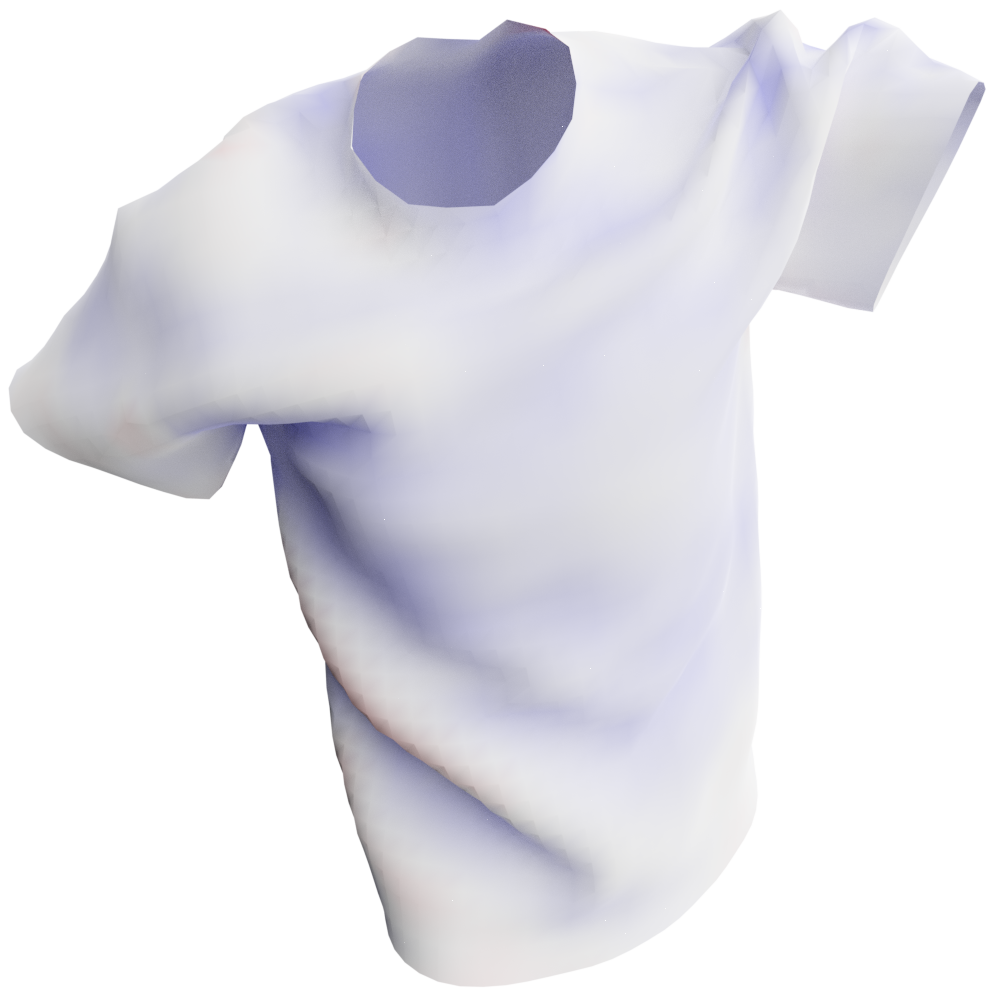}
		\caption{}
		\label{fig:post_buck_untrained}
	\end{subfigure}
	
	\begin{subfigure}[b]{0.32\linewidth}
		\includegraphics[width=\linewidth]{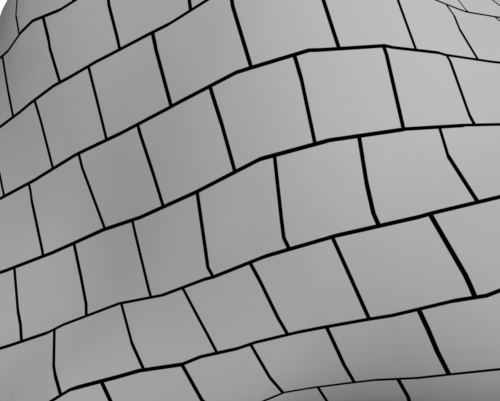}
		\caption{}
		\label{fig:pre_buck_tex}
	\end{subfigure}
	\begin{subfigure}[b]{0.32\linewidth}
		\includegraphics[width=\linewidth]{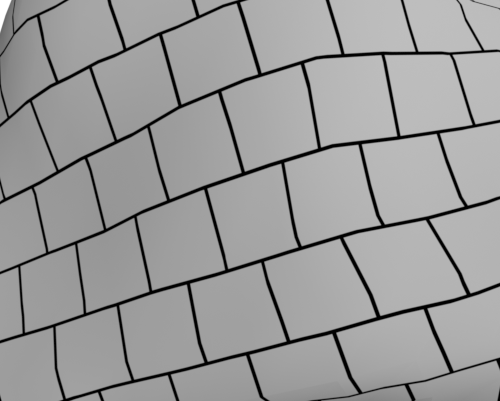}
		\caption{}
		\label{fig:post_ineq_tex2}
	\end{subfigure}
	\begin{subfigure}[b]{0.32\linewidth}
		\includegraphics[width=\linewidth]{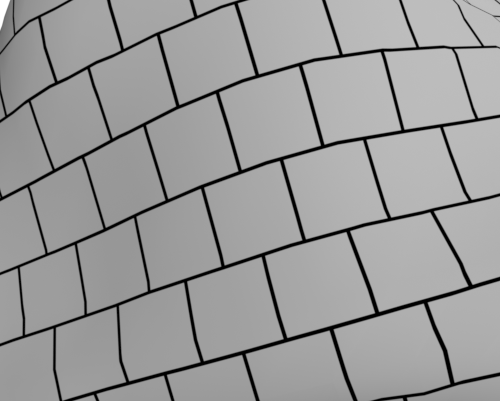}
		\caption{}
		\label{fig:post_buck_trained_tex}
	\end{subfigure}
	\caption{
		(a) Network output depicting over-stretching in red and over-compression in blue (pure red indicates 1.25 times stretching, pure blue indicates 0.75 times compression, and white indicates no distortion). 
		(b) Results obtained after applying the inextensibility postprocess from Section~\ref{sec:inext_post} to (a).
		(c) Results obtained after applying the quasistatic postprocess to (a).
        Sub-figures (d), (e), and (f) show zoomed-in textured views of (a), (b), and (c) respectively, illustrating the removal of non-physical in-plane distortion.
	}
	\label{fig:buckling_img}
	\includegraphics[width=0.45\linewidth]{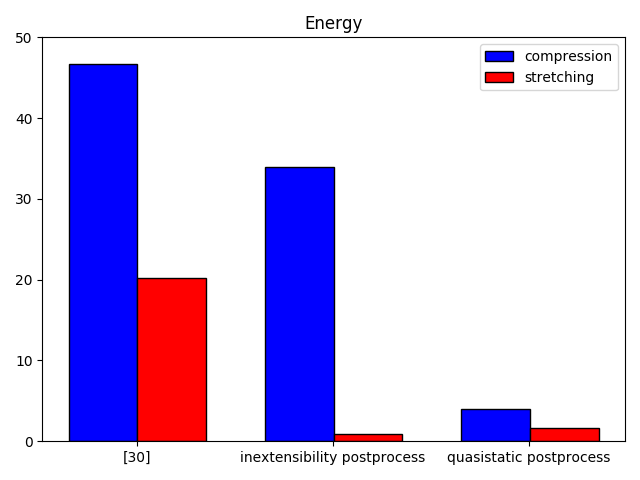}
	\caption{
		The quasistatic postprocess removes spuriously high stretching energy as well as the inextensibility postprocess did, but is also able to remove spurious compression energy, unlike the inextensibility postprocess. The energy is averaged over all the examples in the test set.
	}
	\label{fig:energy_compare}
\end{figure}


Figure~\ref{fig:sec5_springs} illustrates the material model we utilize. For notational clarity, we rename the edges of the triangle mesh from $\E$ to $\E_L$ going forward, and denote the fictional bending edges as $\E_B$ and zero-length axial bending springs as $\E_A$. Then the total spring energy is a combination of $E_{edge}(r)=\sum_{\E_L}\frac{1}{2}k_e^L\left(l_e^L(\sim)-\bar{l}_{e}^L\right)^2$, $E_{bending}(r)=\sum_{\E_B}\frac{1}{2}k_e^B\left(l_e^B(\sim)-\bar{l}_{e}^B\right)^2$, and $E_{axial}(r)=\sum_{\E_A}\frac{1}{2}k_e^A\left((1-u)r_e^a+ur_e^b-(1-v)r_e^c-vr_e^d\right)^2$. 
This total energy is approximated by a truncated Taylor expansion $E(r+dr)\approx E(r)+J^T(r) dr + \frac{1}{2} dr^T H(r) dr$, which is minimized by $H(r)dr=-J(r)$. At each iteration, following a Newton-Raphson approach, we invert $H(r)$ to find a direction $dr$ for a line search aiming to minimize the energy. Since zero-length springs only stretch and do not compress, $H_{axial}$ is symmetric positive semi-definite (SPSD); however, both $H_{edge}$ and $H_{bending}$ suffer from indefiniteness that we remedy via \cite{teran2005robust} so that a fast conjugate gradient solver may be used. As discussed in \cite{teran2005robust}, the fix for definiteness does not adversely impact the final solution to the minimization. Note that Figure~\ref{fig:sec5_springs} depicts a tetrahedron which may invert towards a spurious inside out steady state, see \eg \cite{irving2004invertible}. Although this too can be addressed in our formulation, we omit the discussion since this is not an issue for the flat rest states considered herein. 
Similar to Section~\ref{sec:inext_post}, we add additional zero-length springs connecting $\sim$ to $\net$, with $E_Z=\sum_i \frac{1}{2} k_i^z \lVert \net_i-\sim_i \rVert_2^2$. Then $E_Z$ is added to the total energy minimized, noting that the hessian of $E_Z$ is SPSD. Unlike in Section~\ref{sec:inext_post} where only the relative not the overall scaling of $k_i^z$ mattered, here the zero-length springs compete with the cloth material model to achieve an equilibrium; thus, stronger $k_i^z$ adhere $\sim$ to $\net$ and weaker $k_i^z$ allow the cloth to equilibrate independent of the network prediction. Figure~\ref{fig:stiffness} shows the postprocessed results for various global scaling factors $\mu$. Surprisingly, an adequate value of $k_i^z$ matches the ground truth quite well, even though the entirety of the effects from collisions and gravity are modeled only by the network.  This is a significant optimization as compared to a pure numerical simulation because one does not need costly geometric representations of the collision body, \eg level sets \cite{osher2006level}, or algorithmic acceleration structures to make processing collisions more feasible. 

\subsection{Examples}
Figure~\ref{fig:buckling_img} shows how the quasistatic postprocess greatly improves the results removing non-physical in-plane distortions as compared to both the network results as well as  the results obtained using the inextensibility postprocess. In addition, Figure~\ref{fig:energy_compare} shows that similar improvements are obtained when considering the energies.  

%% file: section6.tex
\section{Buckling and Inextensibility Prior}
\label{sec:buck_prior}
Similar to Section~\ref{sec:inext_prior} we minimize an energy containing terms of the form $\lVert \opt-r_T\rVert_2^2$. Again, the minimization process requires the derivatives of this energy with respect to the network weights $w$, and since the derivatives with respect to $\opt$ are readily computed and $\pd{\net}{w}$ is accessible from \cite{jin2018pixel}, we only need consider $\pd{\opt}{\net}$. 

Conceptually speaking, the goal is to find a set of network parameters $w$ that determine cloth geometry via the network in \cite{jin2018pixel} such that the zero-length springs attached to that network driven cloth geometry drive the quasistatic simulation mesh to well match the training data. \cite{bao2019high} addressed a similar problem where they solved for a set of parameters that determined blendshape muscle geometry such that attached zero-length springs drove their quasistatic simulation mesh to well match a ground truth target. They utilized the approach in \cite{sifakis2005automatic} to evaluate search directions for the optimization (noting that \cite{sifakis2005automatic} solved for muscle activations directly whereas \cite{bao2019high} obtained activations indirectly using zero-length springs attached to kinematically driven geometry as proposed in \cite{cong2016art}).

Although the material model forces only depend on $\opt$ unlike the zero length springs that depend on both $\opt$ and $\net$, $\opt$ depends on $\net$ and so the dependencies may be written as $f_Z(\opt(\net),\net)$ and $f_M(\opt(\net))$. The total derivative of the forces $f=f_Z+f_M$ with respect to the network output $\net$ is
\begin{equation}
\label{eq:sec6_1}
 \pd{f_Z}{\opt}\pd{\opt}{\net}+\pd{f_Z}{\net}+\pd{f_M}{\opt}\pd{\opt}{\net}=\pd{f_Z+f_M}{\opt}\pd{\opt}{\net}+\pd{f_Z}{\net}=\pd{f}{\opt}\pd{\opt}{\net}+\pd{f_Z}{\net}
\end{equation}
implying that the quasistatic net force equal to zero solution may be obtained by solving
\begin{equation}
\label{eq:sec6_2}
 \pd{f}{\opt}\pd{\opt}{\net}=-\pd{f_Z}{\net}
\end{equation}
to find $\pd{\opt}{\net}$.
In addition, noting that force is the negative derivative of potential energy with respect to position, $\pd{f}{\opt}$ is the negative Hessian from Section~\ref{sec:buck_post}. Again using associativity for efficient computation:
\begin{equation}
\label{eq:sec6_3}
 \pd{\Et}{w}=\pd{\Et}{\opt}\pd{\opt}{\net}\pd{\net}{w}=\pd{\Et}{\opt}H^{-1}\pd{f_Z}{\net}\pd{\net}{w}=\left(\pd{\Et}{\opt}H^{-1}\right)\pd{f_Z}{\net}\pd{\net}{w}
\end{equation}
where $\frac{\partial \Et}{\partial \opt}$ is again a row vector. Define the row vector $\eta=\pd{\Et}{\opt}H^{-1}$; then, we may compute $\eta$ by solving $H\eta^T=\pd{\Et}{\opt}^T$ once for each training example. This is the same linear system solved for quasistatics in Section~\ref{sec:buck_post}. Finally, note that $\pd{f_Z}{\net}$ is a sparse matrix and can be computed efficiently. 

\subsection{Examples}
Our cloth consists of 2969 vertices, each with a zero-length spring. The material model has 8787 edge springs, 8661 bending springs, and 8661 axial bending springs. The derivatives of the total energy with respect to the weights $w$ can be computed separately for the terms corresponding to each training example using Equation~\ref{eq:sec6_3}. For each training example, solving the quasistatic problem for $\opt$ takes about $0.07$ seconds, and computing $\pd{E}{\net}$ via $\eta$ takes about $0.01$ seconds.


Similar to Section~\ref{sec:inext_prior_examples}, we train our network on a data set with $1000$ training examples and report the errors in Table~\ref{tab:sec6_1k}. Unlike the inextensibility postprocess, the overall scaling of the zero-length spring stiffness matters for the quasistatic postprocess. Thus, we compare errors for three different global scalings: $\mu=1$, $0.1$, and $0.01$. When $\mu=1$, the quasistatic postprocess reduces SqrtMSE; however, including the postprocess in training generally yields only minor improvements. This is because the zero-length springs are too strong relative to the material model, and thus the material model is unable to provide a significant correction to the output of the network. Table~\ref{tab:sec6_diff_stiff_energy} substantiates this, showing relatively high errors in compression/stretching energies for $\mu=1$. Next, $\mu=0.1$ can be viewed as nearly optimal based on Figure~\ref{fig:stiffness}, and in this case, including the postprocess in training yields up to around $25\%$ improvement. Table~\ref{tab:sec6_diff_stiff_energy} shows that the errors in energy for $\mu=0.1$ are quite small. When the scaling is $0.01$ the springs are too weak, and the postprocess by itself does a poor job matching the ground truth because the material model dominates the network prediction pulling the final results quite far from the ground truth. In this case, including the postprocess in training makes significant improvements.

\begin{table*}[t]
\centering
\begin{minipage}[t]{\textwidth}
\centering
\begin{tabular}{c|c|c|c|c|c|c|c}
\hline
\multirow{2}{*}{Scaling} & \multirow{2}{*}{Method} & \multicolumn{2}{c|}{Training set} & \multicolumn{2}{c|}{Validation set} & \multicolumn{2}{c}{Test set} \\ \cline{3-8}
& & SqrtMSE  & MaxDist  & SqrtMSE & MaxDist & SqrtMSE & MaxDist  \\ \hline
N/A & \cite{jin2018pixel} & $2.5364$ & $10.896$ & $7.5385$ & $29.322$ & $7.5131$ & $29.308$ \\ \hline
\multirow{2}{*}{$\mu = 1$}& Postprocess Only & $2.2014$ & $10.333$ & $7.3195$ & $29.371$ & $7.2918$ & $29.604$ \\ \cline{2-8}
& Trained Postprocess & $2.1754$ & $9.7121$ & $7.2804$ & $29.716$ & $7.2794$ & $31.191$ \\ \hline
\multirow{2}{*}{$\mu = 0.1$}& Postprocess Only & $2.2268$ & $12.881$ & $7.2168$ & $30.371$ & $7.1879$ & $30.599$ \\ \cline{2-8}
& Trained Postprocess & $1.7223$ & $10.864$ & $7.1435$ & $30.246$ & $7.1465$ & $31.191$ \\ \hline
\multirow{2}{*}{$\mu = 0.01$}& Postprocess Only & $4.3621$ & $22.045$ & $8.8966$ & $34.594$ & $8.9365$ & $35.022$ \\ \cline{2-8}
& Trained Postprocess & $2.3120$ & $16.303$ & $7.3002$ & $31.718$ & $7.3596$ & $32.494$ \\ \hline
variable $\mu_i$ & Trained Postprocess & $1.6213$ & $10.969$ & $6.8900$ & $31.027$ & $6.9375$ & $31.816$ \\ \hline
\end{tabular}
\caption{Errors on a data set with 1000 training examples in the energy function. Numbers are in millimeters.}
\label{tab:sec6_1k}
\end{minipage}
\vspace{5mm}

\begin{minipage}{\textwidth}
\centering
\begin{tabular}{c|c|c|c|c|c|c}
\hline
\multirow{2}{*}{Scaling} &  \multicolumn{2}{c|}{Training Set} & \multicolumn{2}{c|}{Validation Set} & \multicolumn{2}{c}{Test Set}  \\ \cline{2-7}
  & Compression & Stretching & Compression & Stretching & Compression & Stretching  \\ \hline
$\mu = 1$ &  $4.5191$ & $1.0058$ & $14.612$ & $4.1526$ & $14.897$ & $4.0208$ \\ \hline
$\mu = 0.1$ &  $0.4444$ & $0.4786$ & $1.7589$ & $0.4661$ & $1.7435$ & $0.4415$ \\ \hline
$\mu = 0.01$ &  $1.5587$ & $0.3319$ & $1.6390$ & $0.6551$ & $1.7534$ & $0.6847$ \\ \hline
variable $\mu_i$ &  $0.3183$ & $0.4580$ & $0.5990$ & $0.4289$ & $0.6433$ & $0.4717$ \\ \hline
\end{tabular}
\caption{Compression/stretching energy errors for different scalings of the zero-length spring stiffness.}
\label{tab:sec6_diff_stiff_energy}
\end{minipage}
\end{table*}

\subsection{Optimizing for Zero-Length Spring Stiffness}
\begin{figure}[t]
\centering
\begin{subfigure}[b]{0.4\linewidth}
	\includegraphics[width=\textwidth]{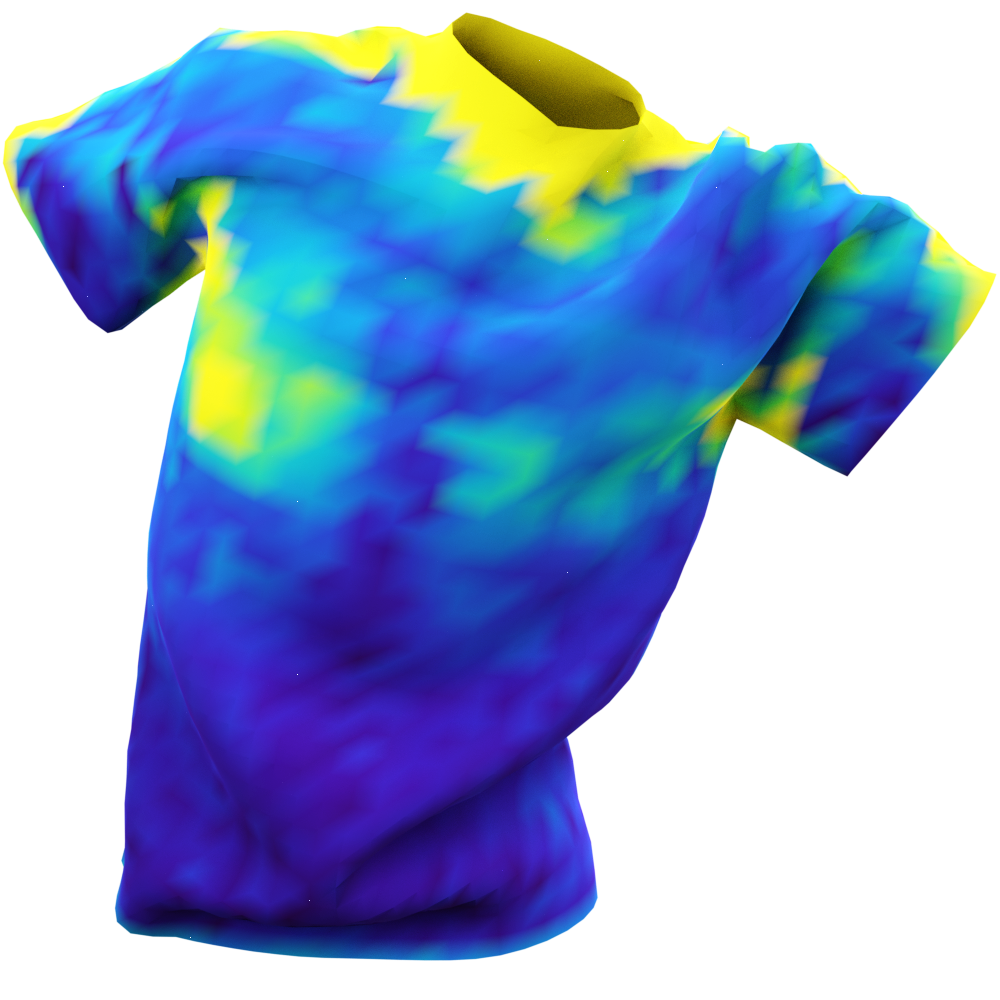}
	\caption{front view}
	\label{fig:sec6_sp_vm_front}
\end{subfigure}
\begin{subfigure}[b]{0.4\linewidth}
    \includegraphics[width=\textwidth]{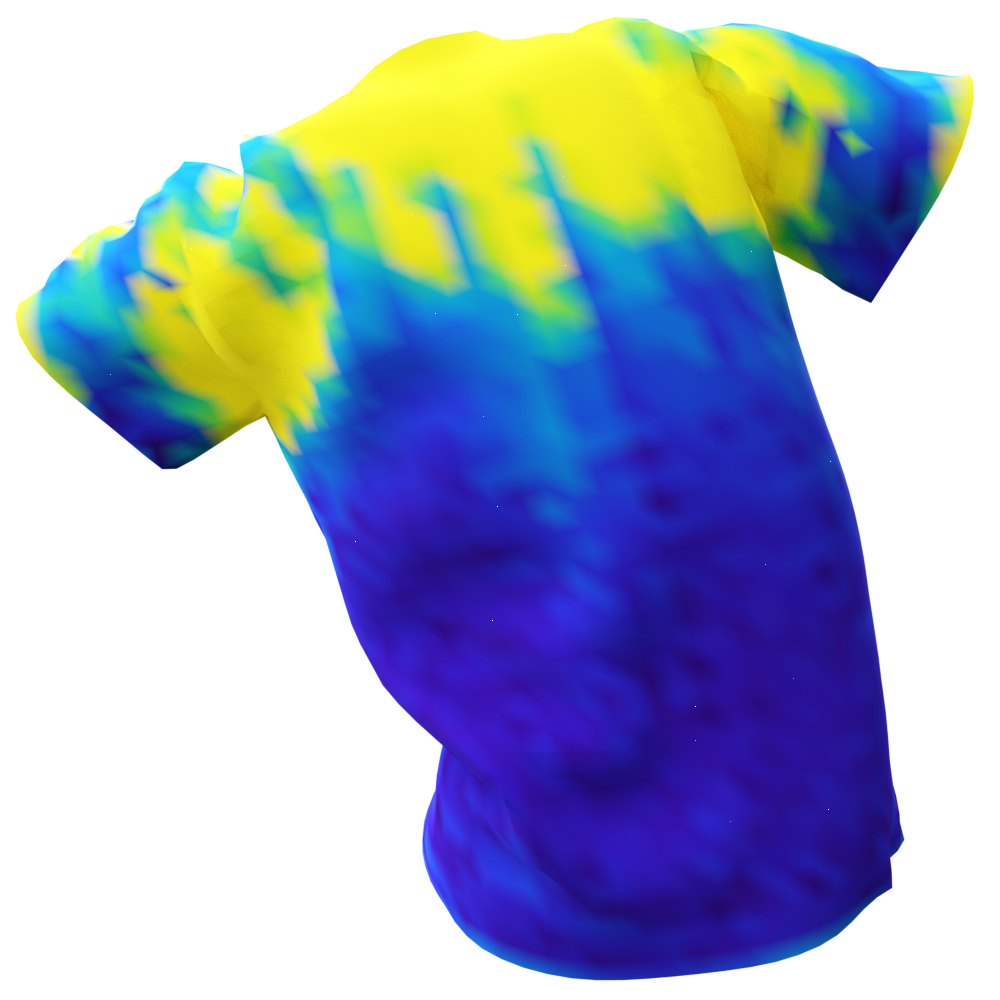}
	\caption{back view}
	\label{fig:sec6_sp_vm_back}
\end{subfigure}
    \caption{Spatially varying scaling. Yellow indicates $\mu_i = 0.2$, and purple indicates $\mu_i = 0.01$. }
    \label{fig:sec6_sp_vm}
\end{figure}

Instead of experimentally determining the best global scaling $\mu$ for the zero-length springs, one could augment the network to choose this parameter itself.
Replacing $\net$ with $\mu$ in Equation~\ref{eq:sec6_1} and the related discussion results in solving
\begin{equation}
\label{eq:sec6_22}
 \pd{f}{\opt}\pd{\opt}{\mu}=-\pd{f_Z}{\mu} 
\end{equation}
to find $\pd{\opt}{\mu}$, similar to Equation~\ref{eq:sec6_2} except with a different right hand side. Then, similar to Equation~\ref{eq:sec6_3}, we obtain:
\begin{equation}
\label{eq:sec6_23}
\pd{\Et}{\mu}=\pd{\Et}{\opt}\pd{\opt}{\mu}=\pd{\Et}{\opt}H^{-1}\pd{f_Z}{\mu}=\left(\pd{\Et}{\opt}H^{-1}\right)\pd{f_Z}{\mu}
\end{equation}
where $\eta=\pd{\Et}{\opt}H^{-1}$ was already computed for Equation~\ref{eq:sec6_3}. Starting with an initial guess of $\mu=0.1$, we obtain $\mu=0.0984$ confirming the optimality of $\mu \approx 0.1$. Furthermore, starting with initial guesses of $\mu=1$ and $\mu=0.01$, the network converged to $\mu=0.1003$ and $0.0903$ respectively. 
Instead of choosing $\mu$ globally, one might ansatz that different regions of the cloth would benefit from different scalings. As an experiment, we trained the network to learn $\mu_i$ on a per-vertex basis using the appropriate analog of Equations~\ref{eq:sec6_22} and \ref{eq:sec6_23}. Figure~\ref{fig:sec6_sp_vm} illustrates that stronger scalings are learned near the upper neck and back as opposed to the midsection of the shirt. As shown in Tables~\ref{tab:sec6_1k} and \ref{tab:sec6_diff_stiff_energy}, allowing the network to determine $\mu_i$ yields quite good results.

%% file: conclusion.tex
\section{Conclusion}
Although networks are trained to match ground truth data and regularized to interpolate reasonably well away from the training data, this is typically in an averaged sense which may allow for poor local behavior and a lack of physical constraints. 
In fact, the resulting mesh could be ill-conditioned or could contain inverted elements, prohibiting any simulation whatsoever. Even when the results may be simulated, the initial conditions may contain large non-physical energies that lead to spurious behaviors including fracture, leading to the failure of the simulator.
Thus, we proposed the notion of a postprocess that projects the output of a network into a feasible state ready for simulation. Notably, this postprocess must be robust enough to accept network output as an initial condition. We also require the postprocess to be differentiable so that we may include it in the network training such that the final result not only has good physical properties but also well matches the training data.

We proposed two such postprocesses in this paper and applied both to the network from \cite{jin2018pixel} that predicts cloth vertex positions from joint angles. The first postprocess simply constrains mesh edges to not overstretch and was reformulated into a second order cone program and shown to greatly reduce the non-physical stretching energy in the network output. The second postprocess was based on a robust quasistaic simulation model that deforms the output of the network into a state with highly improved stretching/compression energies. In both cases, zero length spring constraints were used to tie the network predictions to the final result. In the case of the second order cone program, the final results are independent of the overall scaling/strength of the zero-length spring constraints. In the quasistatics case, the results are sensitive to this scaling, and we showed that the network could be trained to learn the scaling; in fact, it learned a value similar to that hypothesized to be optimal experimentally. Moreover, we also showed that the network could learn a spatially varying scaling that produced quite good results. We hypothesize that even better results could probably be obtained by enforcing various smoothness or regional constraints on the local scaling.

%% file: acknowledgement.tex
	\section*{Acknowledgement}
	
	Research supported in part by ONR N000014-13-1-0346, ONR N00014-17-1-2174, and JD.com. We would like to thank both Reza and Behzad at ONR for supporting our efforts into machine learning. ZG is supported by a VMWare Fellowship. DJ is supported by a Stanford Graduate Fellowship. We would also like to thank Robert Huang and William Tsu for their kind donation of an Nvidia TITAN X GPU which was used to run experiments. This paper is dedicated to the late John McCarthy who coined the term \textit{Artificial Intelligence}; the last author appreciates the many conversations he had with John not only on artificial intelligence but also on fluid dynamics (especially, the mixing of cream into John's coffee).

%% file: appendix.tex
\newpage
\section*{Appendix A}
Here, we provide a column rank analysis for the coefficient matrix in Equation~\ref{eq:inext_differential}.
For the sake of exposition, we write
\begin{equation}
\begin{gathered}
G_0=\begin{bmatrix}
-\sqrt{k_i^z}I & \\
& \ddots
\end{bmatrix},\quad \\
G_e=\begin{bmatrix}
0  \cdots & I & 0  \cdots & -I & 0  \cdots
\end{bmatrix},\quad
I_e=\begin{bmatrix}
0  \cdots & -1 & 0  \cdots \\
\end{bmatrix},\\
\soo=t,\quad \sol=\begin{bmatrix}
\sqrt{k_i^z}(r_i-\hat{r}_i) \\ \vdots
\end{bmatrix},\quad \seo=\lm,\quad \sel=\vec{l}_e(r)
\end{gathered}
\end{equation}
so that Equation~\ref{eq:inext_differential} can be written in more detail as follows:
\begin{equation}
\hspace{-12mm}
\left[
\begin{array}{c:c:c:c}
& \begin{array}{c}
\\ \\ I  
\end{array} & 
\begin{array}{cccc}
\dashed{
\quad & G_0^T \\
-1 & \quad \\
\quad & \quad \\
} &
\dashed{
\quad & G_e^T \\
\quad & \quad \\
I_e^T & \quad \\
} & \cdots
\end{array}
& \\ 
\hdashline
\begin{array}{ccc}
\quad & \quad  & I
\end{array} & & & \\
\hdashline
\begin{array}{c}
\dashed{
\quad & -1 & \quad \\
G_0 & \quad & \quad  
} 
\\
\dashed{
\quad &  \quad  & I_e\\
G_e & \quad & \quad
}
\\ \vdots \\
\end{array}
& & & I \\
\hdashline
& & 
\begin{array}{ccc}
\dashed{
\soo & \sol^T \\
\sol & \soo I 
} & & \\
& \dashed{
\seo & \sel^T \\
\sel & \seo I
} & \\
& & \ddots
\end{array}  
&
\begin{array}{cccc}
\dashed{
\zoo & \zol^T \\
\zol & \zoo I 
} & &  \\
&  \dashed{
\zeo & \zel^T \\
\zel & \zeo I
} & \\
&  & \ddots
\end{array}  
\end{array}
\right]\left[\begin{array}{c}
\vspace{1mm} dr \\ dt \\ d\alpha \vspace{1mm} \\ \hdashline dy \\ \hdashline 
\dashed{
d\zoo \\ d\zol 
} 
\\ \dashed{
d\zeo \\ d\zel
} 
\\ \vdots \\ \hdashline
\dashed{
d\soo \\ d\sol
}
\\ 
\dashed{
d\seo \\ d\sel
} \\ \vdots
\end{array}\right]=\left[\begin{array}{c}
\ \\ \hdashline \ \\ \hdashline dh \\ \hdashline \ 
\end{array}\right].
\end{equation}
The second row partition is $Id\alpha=0$, and thus setting $d\alpha=0$ removes the second row partition as well as the third block column.  The third block row is  $Idy+\sum_e I_e^T d\zeo = 0$, and so $dy$ can be computed from $d\zeo$ independent of the coupled system; thus, the third block row and the fourth block column can be removed. The resulting system has the following condition on any potential null space vector:
\begin{equation}
\label{eq:append_detailed_null_space_system}
\left[
\begin{array}{c:c:c}
& 
\begin{array}{cccc}
\dashed{
& G_0^T \\
-1 & \\
} &
\dashed{
\quad & G_e^T \\
& \\
} & \cdots
\end{array}
& \\ 
\hdashline
\begin{array}{c}
\dashed{
& -1 \\
G_0 & 
}
\\ 
\dashed{
\quad & \quad \\
G_e & \quad
}
\\ \vdots \\
\end{array}
& & I \\
\hdashline
& 
\begin{array}{ccc}
\dashed{
\soo & \sol^T \\
\sol & \soo I 
} & & \\
&  \dashed{
\seo & \sel^T \\
\sel & \seo I
} & \\
&  & \ddots
\end{array}  
&
\begin{array}{ccc}
\dashed{
\zoo & \zol^T \\
\zol & \zoo I 
} & &  \\
&  \dashed{
\zeo & \zel^T \\
\zel & \zeo I
} & \\
& &  \ddots
\end{array}\\
\end{array}
\right]
\left[\begin{array}{c}
\vspace{1mm} \br \\ \bt \vspace{1mm} \\ \hdashline 
\dashed{
\coo \\\col 
} 
\\ \dashed{
\ceo \\ \cel
} 
\\ \vdots \\ \hdashline
\dashed{
\doo \\ \dol
}
\\ 
\dashed{
\deo \\ \del
} \\ \vdots
\end{array}\right]=0.
\end{equation}

From Equation~\ref{eq:sec3:conic_product} , the determinant of $M_q$ is $(q_0^2-q_1^Tq_1)q_0^{m-2}$. When $q_0=0$, then $q_0\geq \lVert q_1\rVert_2$ forces $q_1=0$ and $M_q$ is rank $0$.
When $q_0>0$, $M_q$ has either full rank, or has a single null space vector $\begin{bmatrix}
q_0 \\ -q_1
\end{bmatrix}$. 

Firstly, consider only the first column partition of Equation~\ref{eq:append_detailed_null_space_system}. Since $\begin{bmatrix}
& -1 \\ G_0 &
\end{bmatrix}$ has full rank, the first column partition has full rank amongst itself.

\begin{figure}[t]
\centering
\includegraphics[width=.5\textwidth]{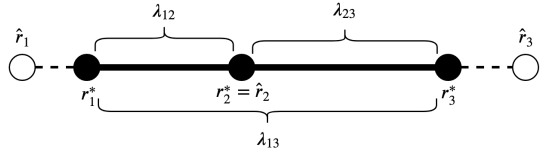}
\caption{Here, the optimal $\opt_1, \opt_2, \opt_3$ are in a linear configuration, targeting the network predicted $\net_1, \net_2$, $\net_3$ also in a linear configuration. All three edges of this triangle need to be maximally stretched in order for it to be a null space candidate. The constraint forces on $\lambda_{12}$ and $\lambda_{23}$ need to balance each other so that $\opt_2$ remains coincident with $\net_2$. The total constraint force of $\lambda_{12}+\lambda_{13}$ on $\opt_1$ needs to remain fixed as well as $\lambda_{23}+\lambda_{13}$ on $\opt_3$. Then, one can freely increase/decrease the force $\lambda_{12}$ and $\lambda_{23}$ applied to $\opt_1$ and $\opt_3$ (keeping the forces on $\opt_2$ balanced), while decreasing/increasing the force on $\lambda_{13}$ so that the net forces on $\opt_1$ and $\opt_3$ do not change.}
\label{fig:appendix_nullspace_example}
\end{figure}

Secondly, expand to consider the first two column partitions of Equation~\ref{eq:append_detailed_null_space_system}. The first two column partitions are block orthogonal, so we only need consider the second column partition. 
If $\soo=0$ (i.e.\ $t$=0), then $\sol=0$ (i.e.\ $r=\hat{r}$, meaning that the input $\hat{r}$ is already optimal, with no over-stretched edges, alleviating the need for any perturbations that would stretch zero-length springs). In this case, with $r=\hat{r}$, the null space vector is unimportant since we simply have $\frac{\partial r}{\partial \hat{r}}=I$. So in the following, we only discuss the case when $\soo>0$.
When $\soo>0$, (i.e.\ $t>0$), we have $t^2=\sum_i k_i^z\lVert r_i- \hat{r}_i\rVert_2^2$ i.e.\ $(\soo)^2-\sol^T\sol=0$. The second block row containing only a $(-1)$ sets $\coo=0$, and then the row with $\soo I$ forces $\col=0$. If $(\seo)^2-\sel^T\sel>0$, then $\begin{bmatrix}
\seo & \sel^T  \\
\sel & \seo I  \\
\end{bmatrix}$ is invertible making $\begin{bmatrix}
\ceo \\ \cel
\end{bmatrix}=0$.  So, only consider the set of all edges where $(\seo)^2-\sel^T\sel=0$ (i.e.\ maximally stretched). First, note that $\begin{bmatrix}
\seo & \sel^T  \\
\sel & \seo I  \\
\end{bmatrix}
\begin{bmatrix}
\ceo \\ \cel
\end{bmatrix}=0$ implies $\begin{bmatrix}
\ceo \\ \cel
\end{bmatrix}=k_e\begin{bmatrix}
\seo \\ -\sel
\end{bmatrix}$ for arbitrary $k_e$. Second, consider $\sum_e G_e^T \cel =0$. The rows for vertex $i$ have the form $\sum_{e\in N_i}\pm k_e\sel=0$ where $N_i$ is the set of maximally stretched edges that contain vertex $i$. A simple example of the null space is shown in Figure~\ref{fig:appendix_nullspace_example}. In general, a null space exists only when degenerate configurations exist within the set of maximally stretched edges where varying constraint forces all produce the same net force on the zero length springs.  

Thirdly, expand to consider the full matrix. We will show that all of the $d$ components of the the null space vector in Equation~\ref{eq:append_detailed_null_space_system} must be zero which means that the null space analysis of the first two column partitions covers the entire null space.

The optimality conditions (Equation~\ref{eq:SOCP_KKT} and $s\circ z=0$) can be written in more detail as follows:

\begin{subequations}
\label{eq:append_detailed_KKT}
\begin{gather}
\left[\begin{array}{c:c:c:c}
& \begin{array}{c}
\\ \\ I  
\end{array} & 
\begin{array}{ccc}
\dashed{
& G_0^T \\
-1 & \\
& \\
} &
\dashed{
& G_e^T \\
& \\
I_e^T & \\
} & \cdots
\end{array}
& \quad \quad \quad \quad \quad \\ 
\hdashline
\begin{array}{ccc}
\quad & \quad  & I
\end{array} & & & \\
\hdashline
\begin{array}{c}
\dashed{
& -1 & \ \\
G_0 & & \  
}\\
\dashed{
& \  & I_e\\
G_e & \ & 
}
\\ \vdots \\
\end{array}
& & & I \\
\end{array}\right]\left[\begin{array}{c}
r\\ t \\ \alpha \\ \hdashline y \\ \hdashline \dashed{
\zoo \\ \zol
} \\ \dashed{
\zeo \\ \zel 
} \\ \vdots \\ \hdashline \dashed{
\soo \\ \sol
}  \\ \dashed{
\seo \\ \sel 
} \\ \vdots
\end{array}\right]=\left[
\begin{array}{c}
\ \\ -1 \\ \\ \hdashline b \\ \hdashline \ \\ G_0\hat{r} \\  \ \\ \ \\ \vdots
\end{array}
\right] \label{eq:append_detailed_KKT_a} \\
\begin{bmatrix}
\soo & \sol^T \\
\sol & \soo I
\end{bmatrix}  \begin{bmatrix}
\zoo \\ \zol
\end{bmatrix}=0,\quad \begin{bmatrix}
\seo & \sel^T \\ 
\sel & \seo I
\end{bmatrix}  \begin{bmatrix}
\zeo \\ \zel
\end{bmatrix}=0.  \label{eq:append_detailed_KKT_b}
\end{gather}
\end{subequations}
The second block row of Equation~\ref{eq:append_detailed_KKT_a} is $(-1)\zoo=-1$, implying $\zoo=1>0$. Switching the first two block columns in Equation~\ref{eq:append_detailed_null_space_system} to make the first column partition strictly diagonal, followed by eliminating the third column partition using the identity matrix results in:
\begin{equation}
\label{eq:append_eliminate_Z}
\left[
\begin{array}{c:c}
& 
\begin{array}{ccc}
\dashed{
-1 & \\
& G_0^T
} &
\dashed{
\quad & \\
& G_e^T
} & \cdots
\end{array} \\ 
\hdashline
\begin{array}{c}
\dashed{
\zoo & -\zol^T G_0 \\
\zol & -\zoo G_0
}\\
\dashed{
\quad & -\zel^T G_e \\
& -\zeo G_e
}\\ \vdots
\end{array} & 
\begin{array}{ccc}
\dashed{
\soo & \sol^T \\
\sol & \soo I 
} & & \\
&  \dashed{
\seo & \sel^T \\
\sel & \seo I
} & \\
&  & \ddots
\end{array}  \\
\end{array}
\right]
\left[\begin{array}{c}
\vspace{1mm} \bt \\ \br \vspace{1mm} \\ \hdashline 
\dashed{
\coo \\ \col}
\\ \dashed{
\ceo \\ \cel
} 
\\ \vdots 
\end{array}\right]=0.
\end{equation}
As a side note, scaling the second column partition of Equation \ref{eq:append_eliminate_Z} by $Z$ is an alternative symmetrization to that we proposed in Equation~\ref{eq:sym_system} (\ie column scaling by $S$). 
When $(\seo)^2-\sel^T\sel>0$, Equation \ref{eq:append_detailed_KKT_b}  implies $\begin{bmatrix}
\zeo \\ \zel
\end{bmatrix}=0$,
which in turn zeros out the corresponding rows in the first column partition of Equation~\ref{eq:append_eliminate_Z}, resulting in
$\begin{bmatrix}
\ceo \\ \cel
\end{bmatrix}=0$ as well. 
Thus, we only need to consider a sub-system with $(\seo)^2-\sel^T\sel=0$ for all $e$ hereafter. 

From  $\seo=\lm>0$ and Equation~\ref{eq:append_detailed_KKT_b}, we have $\begin{bmatrix}
\zeo \\ \zel
\end{bmatrix}=k_e\begin{bmatrix}
\seo \\ -\sel
\end{bmatrix}$ where $k_e\geq 0$ because $\zeo\geq 0$. Substituting this into Equation~\ref{eq:append_eliminate_Z} modifies the corresponding block rows to be
\begin{equation}
\label{eq:append_edge_e}
\begin{bmatrix}
k_e\sel^TG_e & \seo & \sel^T \\
-k_e\seo G_e & \sel & \seo I
\end{bmatrix}\begin{bmatrix}
b_r \\ \ceo \\ \cel
\end{bmatrix}=0.
\end{equation}
Premultiplying with $\begin{bmatrix}
\seo & -\sel^T
\end{bmatrix}$ zeros out the last two columns
obtaining $2k_e\seo\sel^T G_e b_r=0$ or $k_e\sel^TG_e\br=0$ or $\zel^T G_e\br=0$.
The first block row of Equation~\ref{eq:append_eliminate_Z} implies $\coo=0$.
From $\soo>0$, $(\soo)^2-\sol^T\sol=0$, and Equation~\ref{eq:append_detailed_KKT_b}, we have $\begin{bmatrix}
\zoo \\ \zol
\end{bmatrix}=k_0\begin{bmatrix}
\soo \\ -\sol
\end{bmatrix}$ where $k_0>0$ because $\zoo>0$.  Substituting this into Equation~\ref{eq:append_eliminate_Z} modifies the corresponding block rows to be
\begin{equation}
\label{eq:append_row_0}
\begin{bmatrix}
k_0\soo & k_0\sol^TG_0 & \sol^T \\
-k_0\sol & -k_0\soo G_0 & \soo I
\end{bmatrix}\begin{bmatrix}
b_t \\ b_r \\ \col
\end{bmatrix}=0.
\end{equation}
Premultiplying with $\begin{bmatrix}
\soo & \sol^T
\end{bmatrix}$ zeros out the first two columns obtaining $2\soo\sol^T\col=0$ or $\sol^T\col=0$. Then the first row of Equation~\ref{eq:append_row_0} is $k_0\soo \bt+k_0 \sol^T G_0 \br=0$ or $\zoo\bt-\zol^TG_0\br=0$, which can be written as $\zoo\bt+\sum_e \zel^T G_e\br=0$ using the first row of Equation~\ref{eq:append_detailed_KKT_a}. Since we have shown that $\zel=0$ when $(\seo)^2-\sel^T\sel>0$ and $\zel^TG_e\br=0$ when $(\seo)^2-\sel^T\sel=0$, $\sum_e \zel^T G_e\br$ is identically $0$ making $\bt=0$ because $\zoo>0$.

The reduced subsystem becomes:
\begin{equation}
\label{eq:append_trucate_eliminate_Z}
\left[
\begin{array}{c:c}
& 
\begin{array}{ccc}
\hspace{4mm} G_0^T & \hspace{2mm}
\dashed{
\quad & G_e^T
}& \cdots
\end{array} \\ 
\hdashline
\begin{array}{c}
\dashed{
-\zol^T G_0 \\
-\zoo G_0
}\\
\dashed{
-\zel^T G_e \\
-\zeo G_e
}\\ \vdots
\end{array} & 
\begin{array}{ccc}
\dashed{
\sol^T \\
\soo I 
} & & \\
&  \dashed{
\seo & \sel^T \\
\sel & \seo I
} & \\
&  & \ddots
\end{array}  \\
\end{array}
\right]
\left[\begin{array}{c}
\vspace{1mm}  \br \vspace{1mm} \\ \hdashline 
\col
\\ \dashed{
\ceo \\ \cel
} 
\\ \vdots 
\end{array}\right]=0.
\end{equation}
The third block row gives $\col=\frac{\zoo}{\soo}G_0\br$, and the fifth, seventh, \etc block rows give $\cel=\frac{\zeo}{\seo} G_e \br -\frac{\ceo}{\seo} \sel$. Substituting these into the first block row gives $M\br-\sum_e \frac{\ceo}{\seo} G_e^T\sel=0$ where  $M=\left(\frac{\zoo}{\soo}G_0^TG_0 +\sum_e \frac{\zeo}{\seo} G_e^T G_e \right)$ is symmetric positive definite. Thus, Equation~\ref{eq:append_trucate_eliminate_Z} reduces to:
\begin{equation}
\label{eq:append_simplified_system}
\left[
\begin{array}{c:c}
M &  -\frac{1}{\seo} G_e^T\sel \cdots\\
\hdashline
\begin{array}{c}
k_e \sel^T G_e \\
\vdots
\end{array}  & 
\end{array}\right]\left[\begin{array}{c}
\br \\ \hdashline
\ceo \\ \vdots
\end{array}\right]=0
\end{equation}
where the second the the subsequent rows come from premultiplying Equation~\ref{eq:append_edge_e} with $\begin{bmatrix}
\seo & -\sel^T
\end{bmatrix}$. Assuming $k_e\neq 0$, we can row scale by $-\frac{1}{\seo k_e}$ to obtain $\begin{bmatrix}
M & N \\
N^T
\end{bmatrix}\begin{bmatrix}
\br \\ c
\end{bmatrix}=0$ where $N=\begin{bmatrix}
-\frac{1}{\seo} G_e^T\sel \cdots
\end{bmatrix},c=\begin{bmatrix}
\ceo \cdots
\end{bmatrix}^T$. The first equation gives $\br=-M^{-1}Nc$ and the second equation gives $N^T\br=-N^TM^{-1}Nc=0$. Therefore, $c^TN^T M^{-1} Nc=0$ and thus $Nc=0$ and $\br=0$. Finally setting $\br$ and $\bt$ to $0$ eliminates the first column partition in Equation~\ref{eq:append_detailed_null_space_system} so that the identity matrix in the third column partition sets all of the $d$ components of the null space vector to $0$.

Finally, if $k_e=0$ for some edge, then $\begin{bmatrix}
\zeo \\ \zel
\end{bmatrix}=0$, satisfying Equation~\ref{eq:append_detailed_KKT_b} trivially and also removing the columns corresponding to $\begin{bmatrix}
\zeo \\
\zel
\end{bmatrix}$ in Equation~\ref{eq:append_detailed_KKT_a}. Then $\begin{bmatrix}
\seo \\ \sel
\end{bmatrix}$ decouples from the system. That is, such edges need not appear in the coupled optimality conditions, and thus our null space discussion may assume $k_e\neq 0$ with completeness.

%% file: main.bbl
\begin{thebibliography}{56}
\expandafter\ifx\csname natexlab\endcsname\relax\def\natexlab#1{#1}\fi
\providecommand{\url}[1]{\texttt{#1}}
\providecommand{\href}[2]{#2}
\providecommand{\path}[1]{#1}
\providecommand{\DOIprefix}{doi:}
\providecommand{\ArXivprefix}{arXiv:}
\providecommand{\URLprefix}{URL: }
\providecommand{\Pubmedprefix}{pmid:}
\providecommand{\doi}[1]{\href{http://dx.doi.org/#1}{\path{#1}}}
\providecommand{\Pubmed}[1]{\href{pmid:#1}{\path{#1}}}
\providecommand{\bibinfo}[2]{#2}
\ifx\xfnm\relax \def\xfnm[#1]{\unskip,\space#1}\fi
\bibitem[{Ling et~al.(2016)Ling, Jones, and Templeton}]{ling2016machine}
\bibinfo{author}{J.~Ling}, \bibinfo{author}{R.~Jones},
  \bibinfo{author}{J.~Templeton},
\newblock \bibinfo{title}{Machine learning strategies for systems with
  invariance properties},
\newblock \bibinfo{journal}{Journal of Computational Physics}
  \bibinfo{volume}{318} (\bibinfo{year}{2016}) \bibinfo{pages}{22--35}.
\bibitem[{Parish and Duraisamy(2016)}]{parish2016paradigm}
\bibinfo{author}{E.~J. Parish}, \bibinfo{author}{K.~Duraisamy},
\newblock \bibinfo{title}{A paradigm for data-driven predictive modeling using
  field inversion and machine learning},
\newblock \bibinfo{journal}{Journal of Computational Physics}
  \bibinfo{volume}{305} (\bibinfo{year}{2016}) \bibinfo{pages}{758--774}.
\bibitem[{Raissi et~al.(2017)Raissi, Perdikaris, and
  Karniadakis}]{raissi2017machine}
\bibinfo{author}{M.~Raissi}, \bibinfo{author}{P.~Perdikaris},
  \bibinfo{author}{G.~E. Karniadakis},
\newblock \bibinfo{title}{Machine learning of linear differential equations
  using gaussian processes},
\newblock \bibinfo{journal}{Journal of Computational Physics}
  \bibinfo{volume}{348} (\bibinfo{year}{2017}) \bibinfo{pages}{683--693}.
\bibitem[{Raissi and Karniadakis(2018)}]{raissi2018hidden}
\bibinfo{author}{M.~Raissi}, \bibinfo{author}{G.~E. Karniadakis},
\newblock \bibinfo{title}{Hidden physics models: Machine learning of nonlinear
  partial differential equations},
\newblock \bibinfo{journal}{Journal of Computational Physics}
  \bibinfo{volume}{357} (\bibinfo{year}{2018}) \bibinfo{pages}{125--141}.
\bibitem[{Tripathy and Bilionis(2018)}]{tripathy2018deep}
\bibinfo{author}{R.~K. Tripathy}, \bibinfo{author}{I.~Bilionis},
\newblock \bibinfo{title}{Deep uq: Learning deep neural network surrogate
  models for high dimensional uncertainty quantification},
\newblock \bibinfo{journal}{Journal of Computational Physics}
  \bibinfo{volume}{375} (\bibinfo{year}{2018}) \bibinfo{pages}{565--588}.
\bibitem[{Sirignano and Spiliopoulos(2018)}]{sirignano2018dgm}
\bibinfo{author}{J.~Sirignano}, \bibinfo{author}{K.~Spiliopoulos},
\newblock \bibinfo{title}{Dgm: A deep learning algorithm for solving partial
  differential equations},
\newblock \bibinfo{journal}{Journal of Computational Physics}
  \bibinfo{volume}{375} (\bibinfo{year}{2018}) \bibinfo{pages}{1339--1364}.
\bibitem[{Qi et~al.(2019)Qi, Lu, Scardovelli, Zaleski, and
  Tryggvason}]{qi2019computing}
\bibinfo{author}{Y.~Qi}, \bibinfo{author}{J.~Lu},
  \bibinfo{author}{R.~Scardovelli}, \bibinfo{author}{S.~Zaleski},
  \bibinfo{author}{G.~Tryggvason},
\newblock \bibinfo{title}{Computing curvature for volume of fluid methods using
  machine learning},
\newblock \bibinfo{journal}{Journal of Computational Physics}
  \bibinfo{volume}{377} (\bibinfo{year}{2019}) \bibinfo{pages}{155--161}.
\bibitem[{Chang and Zhang(2019)}]{chang2019identification}
\bibinfo{author}{H.~Chang}, \bibinfo{author}{D.~Zhang},
\newblock \bibinfo{title}{Identification of physical processes via combined
  data-driven and data-assimilation methods},
\newblock \bibinfo{journal}{Journal of Computational Physics}
  \bibinfo{volume}{393} (\bibinfo{year}{2019}) \bibinfo{pages}{337--350}.
\bibitem[{Raissi et~al.(2019)Raissi, Perdikaris, and
  Karniadakis}]{raissi2019physics}
\bibinfo{author}{M.~Raissi}, \bibinfo{author}{P.~Perdikaris},
  \bibinfo{author}{G.~E. Karniadakis},
\newblock \bibinfo{title}{Physics-informed neural networks: A deep learning
  framework for solving forward and inverse problems involving nonlinear
  partial differential equations},
\newblock \bibinfo{journal}{Journal of Computational Physics}
  \bibinfo{volume}{378} (\bibinfo{year}{2019}) \bibinfo{pages}{686--707}.
\bibitem[{Yeo and Melnyk(2019)}]{yeo2019deep}
\bibinfo{author}{K.~Yeo}, \bibinfo{author}{I.~Melnyk},
\newblock \bibinfo{title}{Deep learning algorithm for data-driven simulation of
  noisy dynamical system},
\newblock \bibinfo{journal}{Journal of Computational Physics}
  \bibinfo{volume}{376} (\bibinfo{year}{2019}) \bibinfo{pages}{1212--1231}.
\bibitem[{Zhu et~al.(2019)Zhu, Zabaras, Koutsourelakis, and
  Perdikaris}]{zhu2019physics}
\bibinfo{author}{Y.~Zhu}, \bibinfo{author}{N.~Zabaras}, \bibinfo{author}{P.-S.
  Koutsourelakis}, \bibinfo{author}{P.~Perdikaris},
\newblock \bibinfo{title}{Physics-constrained deep learning for
  high-dimensional surrogate modeling and uncertainty quantification without
  labeled data},
\newblock \bibinfo{journal}{Journal of Computational Physics}
  \bibinfo{volume}{394} (\bibinfo{year}{2019}) \bibinfo{pages}{56--81}.
\bibitem[{Gibou et~al.(2019)Gibou, Hyde, and Fedkiw}]{gibou2019sharp}
\bibinfo{author}{F.~Gibou}, \bibinfo{author}{D.~Hyde},
  \bibinfo{author}{R.~Fedkiw},
\newblock \bibinfo{title}{Sharp interface approaches and deep learning
  techniques for multiphase flows},
\newblock \bibinfo{journal}{Journal of Computational Physics}
  \bibinfo{volume}{380} (\bibinfo{year}{2019}) \bibinfo{pages}{442--463}.
\bibitem[{Stinis et~al.(2019)Stinis, Hagge, Tartakovsky, and
  Yeung}]{stinis2019enforcing}
\bibinfo{author}{P.~Stinis}, \bibinfo{author}{T.~Hagge}, \bibinfo{author}{A.~M.
  Tartakovsky}, \bibinfo{author}{E.~Yeung},
\newblock \bibinfo{title}{Enforcing constraints for interpolation and
  extrapolation in generative adversarial networks},
\newblock \bibinfo{journal}{Journal of Computational Physics}
  \bibinfo{volume}{397} (\bibinfo{year}{2019}) \bibinfo{pages}{108844}.
\bibitem[{Goodfellow et~al.(2016)Goodfellow, Bengio, and
  Courville}]{goodfellow2016deep}
\bibinfo{author}{I.~Goodfellow}, \bibinfo{author}{Y.~Bengio},
  \bibinfo{author}{A.~Courville}, \bibinfo{title}{Deep learning},
  \bibinfo{publisher}{MIT press}, \bibinfo{year}{2016}.
\bibitem[{Hastie et~al.(2005)Hastie, Tibshirani, Friedman, and
  Franklin}]{hastie2005elements}
\bibinfo{author}{T.~Hastie}, \bibinfo{author}{R.~Tibshirani},
  \bibinfo{author}{J.~Friedman}, \bibinfo{author}{J.~Franklin},
\newblock \bibinfo{title}{The elements of statistical learning: data mining,
  inference and prediction},
\newblock \bibinfo{journal}{The Mathematical Intelligencer}
  \bibinfo{volume}{27} (\bibinfo{year}{2005}) \bibinfo{pages}{83--85}.
\bibitem[{Ng(2019)}]{andrew2019lecture}
\bibinfo{author}{A.~Ng}, \bibinfo{title}{Cs229 lecture notes on learning
  theory}, \bibinfo{year}{2019}. \URLprefix
  \url{http://cs229.stanford.edu/summer2019/cs229-notes4.pdf}.
\bibitem[{Li et~al.(2019)Li, Johnson, and Yeung}]{li2019lecture}
\bibinfo{author}{F.-F. Li}, \bibinfo{author}{J.~Johnson},
  \bibinfo{author}{S.~Yeung}, \bibinfo{title}{Cs231n lecture notes on training
  neural networks}, \bibinfo{year}{2019}. \URLprefix
  \url{http://cs231n.stanford.edu/slides/2019/cs231n_2019_lecture08.pdf}.
\bibitem[{Broyden(1970)}]{broyden1970convergence}
\bibinfo{author}{C.~G. Broyden},
\newblock \bibinfo{title}{The convergence of a class of double-rank
  minimization algorithms 1. general considerations},
\newblock \bibinfo{journal}{IMA Journal of Applied Mathematics}
  \bibinfo{volume}{6} (\bibinfo{year}{1970}) \bibinfo{pages}{76--90}.
\bibitem[{Liu and Nocedal(1989)}]{liu1989limited}
\bibinfo{author}{D.~C. Liu}, \bibinfo{author}{J.~Nocedal},
\newblock \bibinfo{title}{On the limited memory bfgs method for large scale
  optimization},
\newblock \bibinfo{journal}{Mathematical programming} \bibinfo{volume}{45}
  (\bibinfo{year}{1989}) \bibinfo{pages}{503--528}.
\bibitem[{Dean et~al.(2012)Dean, Corrado, Monga, Chen, Devin, Le, Mao, Ranzato,
  Senior, Tucker, Yang, and Ng}]{dean2012large}
\bibinfo{author}{J.~Dean}, \bibinfo{author}{G.~S. Corrado},
  \bibinfo{author}{R.~Monga}, \bibinfo{author}{K.~Chen},
  \bibinfo{author}{M.~Devin}, \bibinfo{author}{Q.~V. Le},
  \bibinfo{author}{M.~Z. Mao}, \bibinfo{author}{M.~Ranzato},
  \bibinfo{author}{A.~Senior}, \bibinfo{author}{P.~Tucker},
  \bibinfo{author}{K.~Yang}, \bibinfo{author}{A.~Y. Ng},
\newblock \bibinfo{title}{Large scale distributed deep networks},
\newblock in: \bibinfo{booktitle}{Proceedings of the 25th International
  Conference on Neural Information Processing Systems - Volume 1}, NIPS'12,
  \bibinfo{publisher}{Curran Associates Inc.}, \bibinfo{address}{USA},
  \bibinfo{year}{2012}, pp. \bibinfo{pages}{1223--1231}.
\bibitem[{Robbins and Monro(1951)}]{robbins1951stochastic}
\bibinfo{author}{H.~Robbins}, \bibinfo{author}{S.~Monro},
\newblock \bibinfo{title}{A stochastic approximation method},
\newblock \bibinfo{journal}{The Annals of Mathematical Statistics}
  (\bibinfo{year}{1951}) \bibinfo{pages}{400--407}.
\bibitem[{Wright(2015)}]{wright2015coordinate}
\bibinfo{author}{S.~J. Wright},
\newblock \bibinfo{title}{Coordinate descent algorithms},
\newblock \bibinfo{journal}{Mathematical Programming} \bibinfo{volume}{151}
  (\bibinfo{year}{2015}) \bibinfo{pages}{3--34}.
\bibitem[{Bao et~al.(2018)Bao, Hua, and Fedkiw}]{bao2018improved}
\bibinfo{author}{M.~Bao}, \bibinfo{author}{X.~Hua},
  \bibinfo{author}{R.~Fedkiw},
\newblock \bibinfo{title}{Improved search strategies for determining facial
  expression},
\newblock \bibinfo{journal}{arXiv preprint arXiv:1812.02897}
  (\bibinfo{year}{2018}).
\bibitem[{Polyak(1964)}]{polyak1964some}
\bibinfo{author}{B.~T. Polyak},
\newblock \bibinfo{title}{Some methods of speeding up the convergence of
  iteration methods},
\newblock \bibinfo{journal}{USSR Computational Mathematics and Mathematical
  Physics} \bibinfo{volume}{4} (\bibinfo{year}{1964}) \bibinfo{pages}{1--17}.
\bibitem[{Nesterov(1983)}]{nesterov1983method}
\bibinfo{author}{Y.~E. Nesterov},
\newblock \bibinfo{title}{A method for solving the convex programming problem
  with convergence rate $o(1/k^2)$},
\newblock in: \bibinfo{booktitle}{Doklady Akademii nauk SSSR}, volume
  \bibinfo{volume}{269}, \bibinfo{year}{1983}, pp. \bibinfo{pages}{543--547}.
\bibitem[{Duchi et~al.(2011)Duchi, Hazan, and Singer}]{duchi2011adaptive}
\bibinfo{author}{J.~Duchi}, \bibinfo{author}{E.~Hazan},
  \bibinfo{author}{Y.~Singer},
\newblock \bibinfo{title}{Adaptive subgradient methods for online learning and
  stochastic optimization},
\newblock \bibinfo{journal}{Journal of Machine Learning Research}
  \bibinfo{volume}{12} (\bibinfo{year}{2011}) \bibinfo{pages}{2121--2159}.
\bibitem[{Zeiler(2012)}]{zeiler2012adadelta}
\bibinfo{author}{M.~D. Zeiler},
\newblock \bibinfo{title}{Adadelta: an adaptive learning rate method},
\newblock \bibinfo{journal}{arXiv preprint arXiv:1212.5701}
  (\bibinfo{year}{2012}).
\bibitem[{Tieleman and Hinton(2012)}]{tieleman2012lecture}
\bibinfo{author}{T.~Tieleman}, \bibinfo{author}{G.~Hinton},
\newblock \bibinfo{title}{Lecture 6.5-rmsprop: Divide the gradient by a running
  average of its recent magnitude},
\newblock \bibinfo{journal}{COURSERA: Neural networks for machine learning}
  \bibinfo{volume}{4} (\bibinfo{year}{2012}) \bibinfo{pages}{26--31}.
\bibitem[{Kingma and Ba(2014)}]{kingma2014adam}
\bibinfo{author}{D.~P. Kingma}, \bibinfo{author}{J.~Ba},
\newblock \bibinfo{title}{Adam: A method for stochastic optimization},
\newblock \bibinfo{journal}{arXiv preprint arXiv:1412.6980}
  (\bibinfo{year}{2014}).
\bibitem[{Dozat(2016)}]{dozat2016incorporating}
\bibinfo{author}{T.~Dozat},
\newblock \bibinfo{title}{Incorporating nesterov momentum into adam}
  (\bibinfo{year}{2016}).
\bibitem[{{Deng} et~al.(2009){Deng}, {Dong}, {Socher}, {Li}, {Kai Li}, and {Li
  Fei-Fei}}]{imagenet_cvpr09}
\bibinfo{author}{J.~{Deng}}, \bibinfo{author}{W.~{Dong}},
  \bibinfo{author}{R.~{Socher}}, \bibinfo{author}{L.~{Li}},
  \bibinfo{author}{{Kai Li}}, \bibinfo{author}{{Li Fei-Fei}},
\newblock \bibinfo{title}{Imagenet: A large-scale hierarchical image database},
\newblock in: \bibinfo{booktitle}{2009 IEEE Conference on Computer Vision and
  Pattern Recognition}, \bibinfo{year}{2009}, pp. \bibinfo{pages}{248--255}.
\bibitem[{Jin et~al.(2018)Jin, Zhu, Geng, and Fedkiw}]{jin2018pixel}
\bibinfo{author}{N.~Jin}, \bibinfo{author}{Y.~Zhu}, \bibinfo{author}{Z.~Geng},
  \bibinfo{author}{R.~Fedkiw},
\newblock \bibinfo{title}{A pixel-based framework for data-driven clothing},
\newblock \bibinfo{journal}{arXiv preprint arXiv:1812.01677}
  (\bibinfo{year}{2018}).
\bibitem[{Amos and Kolter(2017)}]{amos2017optnet}
\bibinfo{author}{B.~Amos}, \bibinfo{author}{J.~Z. Kolter},
\newblock \bibinfo{title}{Optnet: Differentiable optimization as a layer in
  neural networks},
\newblock in: \bibinfo{booktitle}{Proceedings of the 34th International
  Conference on Machine Learning-Volume 70}, \bibinfo{organization}{JMLR. org},
  \bibinfo{year}{2017}, pp. \bibinfo{pages}{136--145}.
\bibitem[{Boyd and Vandenberghe(2004)}]{boyd2004convex}
\bibinfo{author}{S.~Boyd}, \bibinfo{author}{L.~Vandenberghe},
  \bibinfo{title}{Convex optimization}, \bibinfo{publisher}{Cambridge
  university press}, \bibinfo{year}{2004}.
\bibitem[{Agrawal et~al.(2019)Agrawal, Barratt, Boyd, Busseti, and
  Moursi}]{akshay2019}
\bibinfo{author}{A.~Agrawal}, \bibinfo{author}{S.~Barratt},
  \bibinfo{author}{S.~Boyd}, \bibinfo{author}{E.~Busseti},
  \bibinfo{author}{W.~Moursi},
\newblock \bibinfo{title}{Differentiating through a cone program},
\newblock \bibinfo{journal}{Journal of Applied and Numerical Optimization}
  \bibinfo{volume}{1} (\bibinfo{year}{2019}) \bibinfo{pages}{107--115}.
\bibitem[{Teran et~al.(2003)Teran, Blemker, Hing, and Fedkiw}]{teran2003finite}
\bibinfo{author}{J.~Teran}, \bibinfo{author}{S.~Blemker},
  \bibinfo{author}{V.~Hing}, \bibinfo{author}{R.~Fedkiw},
\newblock \bibinfo{title}{Finite volume methods for the simulation of skeletal
  muscle},
\newblock in: \bibinfo{booktitle}{Proceedings of the 2003 ACM
  SIGGRAPH/Eurographics Symposium on Computer Animation},
  \bibinfo{organization}{Eurographics Association}, \bibinfo{year}{2003}, pp.
  \bibinfo{pages}{68--74}.
\bibitem[{Irving et~al.(2004)Irving, Teran, and Fedkiw}]{irving2004invertible}
\bibinfo{author}{G.~Irving}, \bibinfo{author}{J.~Teran},
  \bibinfo{author}{R.~Fedkiw},
\newblock \bibinfo{title}{Invertible finite elements for robust simulation of
  large deformation},
\newblock in: \bibinfo{booktitle}{Proceedings of the 2004 ACM
  SIGGRAPH/Eurographics Symposium on Computer Animation}, SCA '04,
  \bibinfo{publisher}{Eurographics Association}, \bibinfo{address}{Goslar
  Germany, Germany}, \bibinfo{year}{2004}, pp. \bibinfo{pages}{131--140}.
\bibitem[{Sifakis et~al.(2005)Sifakis, Neverov, and
  Fedkiw}]{sifakis2005automatic}
\bibinfo{author}{E.~Sifakis}, \bibinfo{author}{I.~Neverov},
  \bibinfo{author}{R.~Fedkiw},
\newblock \bibinfo{title}{Automatic determination of facial muscle activations
  from sparse motion capture marker data},
\newblock \bibinfo{journal}{ACM Trans. Graph.} \bibinfo{volume}{24}
  (\bibinfo{year}{2005}) \bibinfo{pages}{417--425}.
\bibitem[{Cong et~al.(2016)Cong, Bhat, and Fedkiw}]{cong2016art}
\bibinfo{author}{M.~Cong}, \bibinfo{author}{K.~S. Bhat},
  \bibinfo{author}{R.~Fedkiw},
\newblock \bibinfo{title}{Art-directed muscle simulation for high-end facial
  animation},
\newblock in: \bibinfo{booktitle}{Proceedings of the ACM SIGGRAPH/Eurographics
  Symposium on Computer Animation}, SCA '16, \bibinfo{publisher}{Eurographics
  Association}, \bibinfo{address}{Goslar Germany, Germany},
  \bibinfo{year}{2016}, pp. \bibinfo{pages}{119--127}.
\bibitem[{Bao et~al.(2019)Bao, Cong, Grabli, and Fedkiw}]{bao2019high}
\bibinfo{author}{M.~Bao}, \bibinfo{author}{M.~Cong},
  \bibinfo{author}{S.~Grabli}, \bibinfo{author}{R.~Fedkiw},
\newblock \bibinfo{title}{High-quality face capture using anatomical muscles},
\newblock in: \bibinfo{booktitle}{Proceedings of the IEEE Conference on
  Computer Vision and Pattern Recognition}, \bibinfo{year}{2019}, pp.
  \bibinfo{pages}{10802--10811}.
\bibitem[{Magnenat-Thalmann et~al.(1988)Magnenat-Thalmann, Laperri\`{e}re, and
  Thalmann}]{Magnenat-Thalmann:1989}
\bibinfo{author}{N.~Magnenat-Thalmann}, \bibinfo{author}{R.~Laperri\`{e}re},
  \bibinfo{author}{D.~Thalmann},
\newblock \bibinfo{title}{Joint-dependent local deformations for hand animation
  and object grasping},
\newblock in: \bibinfo{booktitle}{Proceedings on Graphics Interface '88},
  \bibinfo{publisher}{Canadian Information Processing Society},
  \bibinfo{address}{Toronto, Ont., Canada, Canada}, \bibinfo{year}{1988}, pp.
  \bibinfo{pages}{26--33}.
\bibitem[{Blemker and Delp(2005)}]{Blemker2005}
\bibinfo{author}{S.~S. Blemker}, \bibinfo{author}{S.~L. Delp},
\newblock \bibinfo{title}{Three-dimensional representation of complex muscle
  architectures and geometries},
\newblock \bibinfo{journal}{Annals of Biomedical Engineering}
  \bibinfo{volume}{33} (\bibinfo{year}{2005}) \bibinfo{pages}{661--673}.
\bibitem[{Kavan et~al.(2007)Kavan, Collins, \v{Z}\'{a}ra, and
  O'Sullivan}]{Kavan:2007}
\bibinfo{author}{L.~Kavan}, \bibinfo{author}{S.~Collins},
  \bibinfo{author}{J.~\v{Z}\'{a}ra}, \bibinfo{author}{C.~O'Sullivan},
\newblock \bibinfo{title}{Skinning with dual quaternions},
\newblock in: \bibinfo{booktitle}{Proceedings of the 2007 Symposium on
  Interactive 3D Graphics and Games}, I3D '07, \bibinfo{publisher}{ACM},
  \bibinfo{address}{New York, NY, USA}, \bibinfo{year}{2007}, pp.
  \bibinfo{pages}{39--46}.
\bibitem[{Stavness et~al.(2014)Stavness, S\'{a}nchez, Lloyd, Ho, Wang, Fels,
  and Huang}]{Stavness:2014}
\bibinfo{author}{I.~Stavness}, \bibinfo{author}{C.~A. S\'{a}nchez},
  \bibinfo{author}{J.~Lloyd}, \bibinfo{author}{A.~Ho},
  \bibinfo{author}{J.~Wang}, \bibinfo{author}{S.~Fels},
  \bibinfo{author}{D.~Huang},
\newblock \bibinfo{title}{Unified skinning of rigid and deformable models for
  anatomical simulations},
\newblock in: \bibinfo{booktitle}{SIGGRAPH Asia 2014 Technical Briefs}, SA '14,
  \bibinfo{publisher}{ACM}, \bibinfo{address}{New York, NY, USA},
  \bibinfo{year}{2014}, pp. \bibinfo{pages}{9:1--9:4}.
\bibitem[{Le and Hodgins(2016)}]{Le:2016}
\bibinfo{author}{B.~H. Le}, \bibinfo{author}{J.~K. Hodgins},
\newblock \bibinfo{title}{Real-time skeletal skinning with optimized centers of
  rotation},
\newblock \bibinfo{journal}{ACM Trans. Graph.} \bibinfo{volume}{35}
  (\bibinfo{year}{2016}) \bibinfo{pages}{37:1--37:10}.
\bibitem[{Lee et~al.(2019)Lee, Park, Lee, and Lee}]{Lee:2019}
\bibinfo{author}{S.~Lee}, \bibinfo{author}{M.~Park}, \bibinfo{author}{K.~Lee},
  \bibinfo{author}{J.~Lee},
\newblock \bibinfo{title}{Scalable muscle-actuated human simulation and
  control},
\newblock \bibinfo{journal}{ACM Trans. Graph.} \bibinfo{volume}{38}
  (\bibinfo{year}{2019}) \bibinfo{pages}{73:1--73:13}.
\bibitem[{Le and Lewis(2019)}]{Le:2019}
\bibinfo{author}{B.~H. Le}, \bibinfo{author}{J.~P. Lewis},
\newblock \bibinfo{title}{Direct delta mush skinning and variants},
\newblock \bibinfo{journal}{ACM Trans. Graph.} \bibinfo{volume}{38}
  (\bibinfo{year}{2019}) \bibinfo{pages}{113:1--113:13}.
\bibitem[{Osher and Fedkiw(2006)}]{osher2006level}
\bibinfo{author}{S.~Osher}, \bibinfo{author}{R.~Fedkiw}, \bibinfo{title}{Level
  set methods and dynamic implicit surfaces}, volume \bibinfo{volume}{153},
  \bibinfo{publisher}{Springer Science \& Business Media},
  \bibinfo{year}{2006}.
\bibitem[{Bridson et~al.(2003)Bridson, Marino, and Fedkiw}]{Bridson:2003}
\bibinfo{author}{R.~Bridson}, \bibinfo{author}{S.~Marino},
  \bibinfo{author}{R.~Fedkiw},
\newblock \bibinfo{title}{Simulation of clothing with folds and wrinkles},
\newblock in: \bibinfo{booktitle}{Proceedings of the 2003 ACM
  SIGGRAPH/Eurographics Symposium on Computer Animation}, SCA '03,
  \bibinfo{publisher}{Eurographics Association},
  \bibinfo{address}{Aire-la-Ville, Switzerland, Switzerland},
  \bibinfo{year}{2003}, pp. \bibinfo{pages}{28--36}.
\bibitem[{Bridson et~al.(2002)Bridson, Fedkiw, and
  Anderson}]{bridson2002robust}
\bibinfo{author}{R.~Bridson}, \bibinfo{author}{R.~Fedkiw},
  \bibinfo{author}{J.~Anderson},
\newblock \bibinfo{title}{Robust treatment of collisions, contact and friction
  for cloth animation},
\newblock in: \bibinfo{booktitle}{Proceedings of the 29th Annual Conference on
  Computer Graphics and Interactive Techniques}, SIGGRAPH '02,
  \bibinfo{publisher}{ACM}, \bibinfo{address}{New York, NY, USA},
  \bibinfo{year}{2002}, pp. \bibinfo{pages}{594--603}.
\bibitem[{Selle et~al.(2009)Selle, Su, Irving, and Fedkiw}]{selle2008robust}
\bibinfo{author}{A.~Selle}, \bibinfo{author}{J.~Su},
  \bibinfo{author}{G.~Irving}, \bibinfo{author}{R.~Fedkiw},
\newblock \bibinfo{title}{Robust high-resolution cloth using parallelism,
  history-based collisions, and accurate friction},
\newblock \bibinfo{journal}{IEEE Transactions on Visualization and Computer
  Graphics} \bibinfo{volume}{15} (\bibinfo{year}{2009})
  \bibinfo{pages}{339--350}.
\bibitem[{Jin et~al.(2017)Jin, Lu, Geng, and Fedkiw}]{jin2017inequality}
\bibinfo{author}{N.~Jin}, \bibinfo{author}{W.~Lu}, \bibinfo{author}{Z.~Geng},
  \bibinfo{author}{R.~Fedkiw},
\newblock \bibinfo{title}{Inequality cloth},
\newblock in: \bibinfo{booktitle}{Proceedings of the ACM SIGGRAPH /
  Eurographics Symposium on Computer Animation}, SCA '17,
  \bibinfo{publisher}{ACM}, \bibinfo{address}{New York, NY, USA},
  \bibinfo{year}{2017}, pp. \bibinfo{pages}{16:1--16:10}.
\bibitem[{Hughes et~al.(1977)Hughes, Taylor, and
  Kanoknukulchai}]{taylor1977simple}
\bibinfo{author}{T.~Hughes}, \bibinfo{author}{R.~Taylor},
  \bibinfo{author}{W.~Kanoknukulchai},
\newblock \bibinfo{title}{A simple and efficient finite element for plate
  bending},
\newblock \bibinfo{journal}{International Journal for Numerical Methods in
  Engineering} \bibinfo{volume}{11} (\bibinfo{year}{1977})
  \bibinfo{pages}{1529--1543}.
\bibitem[{Domahidi et~al.(2013)Domahidi, Chu, and Boyd}]{domahidi2013ecos}
\bibinfo{author}{A.~Domahidi}, \bibinfo{author}{E.~Chu},
  \bibinfo{author}{S.~Boyd},
\newblock \bibinfo{title}{{ECOS}: An {SOCP} solver for embedded systems},
\newblock in: \bibinfo{booktitle}{2013 European Control Conference (ECC)},
  \bibinfo{organization}{IEEE}, \bibinfo{year}{2013}, pp.
  \bibinfo{pages}{3071--3076}.
\bibitem[{Selle et~al.(2008)Selle, Lentine, and Fedkiw}]{selle2008mass}
\bibinfo{author}{A.~Selle}, \bibinfo{author}{M.~Lentine},
  \bibinfo{author}{R.~Fedkiw},
\newblock \bibinfo{title}{A mass spring model for hair simulation},
\newblock \bibinfo{journal}{ACM Trans. Graph.} \bibinfo{volume}{27}
  (\bibinfo{year}{2008}) \bibinfo{pages}{64:1--64:11}.
\bibitem[{Teran et~al.(2005)Teran, Sifakis, Irving, and
  Fedkiw}]{teran2005robust}
\bibinfo{author}{J.~Teran}, \bibinfo{author}{E.~Sifakis},
  \bibinfo{author}{G.~Irving}, \bibinfo{author}{R.~Fedkiw},
\newblock \bibinfo{title}{Robust quasistatic finite elements and flesh
  simulation},
\newblock in: \bibinfo{booktitle}{Proceedings of the 2005 ACM
  SIGGRAPH/Eurographics Symposium on Computer Animation}, SCA '05,
  \bibinfo{publisher}{ACM}, \bibinfo{address}{New York, NY, USA},
  \bibinfo{year}{2005}, pp. \bibinfo{pages}{181--190}.

\end{thebibliography}
